\documentclass[conference]{IEEEtran}
\IEEEoverridecommandlockouts
\usepackage{cite}
\usepackage{amsmath,amssymb,amsfonts}
\usepackage{algorithmic}
\usepackage{algorithm}

\usepackage{graphicx}
\usepackage{subfigure}
\usepackage{textcomp}
\usepackage{xcolor}

\usepackage{url}
\usepackage{verbatim} 
\usepackage{multirow}
\usepackage{mdwlist}
\usepackage[export]{adjustbox}

\def\BibTeX{{\rm B\kern-.05em{\sc i\kern-.025em b}\kern-.08em
    T\kern-.1667em\lower.7ex\hbox{E}\kern-.125emX}}
    
\begin{document}
\title{IAAT: A Input-Aware Adaptive Tuning framework for Small GEMM}

\author{\IEEEauthorblockN{Jianyu Yao\IEEEauthorrefmark{1}\IEEEauthorrefmark{2},
Boqian Shi\IEEEauthorrefmark{3},
Chunyang Xiang\IEEEauthorrefmark{1}\IEEEauthorrefmark{2}\IEEEauthorrefmark{4}, 
Haipeng Jia\IEEEauthorrefmark{1}\IEEEauthorrefmark{4},
Chendi Li\IEEEauthorrefmark{1}\IEEEauthorrefmark{2},
Hang Cao\IEEEauthorrefmark{1}\IEEEauthorrefmark{2},
Yunquan Zhang\IEEEauthorrefmark{1}}

\IEEEauthorblockA{\IEEEauthorrefmark{1}SKL of Computer Architecture, Institute of Computing Technology, Chinese Academy of Sciences, Beijing, China}
\IEEEauthorblockA{\IEEEauthorrefmark{2}University of Chinese Academy of Sciences, Beijing, China}
\IEEEauthorblockA{\IEEEauthorrefmark{3}Indiana University Bloomington, Indiana, USA}
\IEEEauthorblockA{\IEEEauthorrefmark{4}Corresponding Author}
\IEEEauthorblockA{\{yaojianyu19f, xiangchunyang, jiahaipeng, lichendi19s, caohang, zyq\}@ict.ac.cn, shiboqi@iu.edu}}
\maketitle

\begin{abstract}
GEMM with the small size of input matrices is becoming widely used in many fields like HPC and machine learning. Although many famous BLAS libraries already supported small GEMM, they cannot achieve near-optimal performance. This is because the costs of pack operations are high and frequent boundary processing cannot be neglected. This paper proposes an input-aware adaptive tuning framework(IAAT) for small GEMM to overcome the performance bottlenecks in state-of-the-art implementations. IAAT consists of two stages, the install-time stage and the run-time stage. In the run-time stage, IAAT tiles matrices into blocks to alleviate boundary processing. This stage utilizes an input-aware adaptive tile algorithm and plays the role of runtime tuning. In the install-time stage, IAAT auto-generates hundreds of kernels of different sizes to remove pack operations. Finally, IAAT finishes the computation of small GEMM by invoking different kernels, which corresponds to the size of blocks. The experimental results show that IAAT gains better performance than other BLAS libraries on ARMv8 platform.
\end{abstract}


\section{Introduction}
\label{sec:introduction}
General matrix multiplication(GEMM), as one of the most important numerical algorithm in dense linear algebra, has been exhaustively studied over the years\cite{goto2008anatomy, wang2013augem, van2015blis}. Many famous BLAS(Basic Linear Algebra Subprograms) libraries, like Intel MKL\cite{intel}, OpenBLAS\cite{wang2013augem}, BLIS\cite{van2015blis}, and ARMPL\cite{armpl}, already implemented high-performance GEMM. GEMM is used to compute $C=\alpha A\times B + \beta C$. Here $C$, $A$, $B$ are $M\times N$, $M\times K$, and $K \times N$ matrices, respectively.

In recent years, small GEMM becomes more and more important in many fields, such as machine learning\cite{hinton2018matrix}, sparse matrix\cite{borvstnik2014sparse}, and fluid dynamics\cite{wozniak2016gimmik}. Many CNNs algorithms use the small matrix on their fully connected layers\cite{CNNinFClayer, wang2021high}. 
Caffe \cite{caffe} is a famous deep learning framework. 
Comparing to BLIS that has not optimized small GEMM, the performance of Caffe utilized optimized BLIS can obtain a performance improvement of 17\%\cite{2017AcceleratingMLbyBLIS}.
By using the optimized implementation of small GEMM, Geoffrey Hinton reduced the number of parameters by a factor of 15 to 310K compared with their baseline CNN with 4.2M parameters \cite{hinton2018matrix}.
In this paper, we define small GEMM as, $\sqrt[3]{MNK}\le 80$, when transposition of input matrices is not TN(TN will be explained in TABLE \ref{tab:ALL GENERATED KERNELS}), or $\sqrt[3]{MNK}\le 32$ when transposition of input matrices is TN. This definition will be explained in Section \ref{sec:performance}.

Traditional implementation and optimization methods of GEMM mainly have three steps: block step, pack step and compute step. Block step tiles matrices into a series of small blocks based on features of the hardware, e.g., TLB, size of the L2 cache. Pack step packs these small blocks based on kernel size to ensure continuity of memory access during kernel calculation. Compute step uses one high-performance kernel with boundary processing to compute matrix multiplication. Because input matrices are relatively large, pack step can massively reduce cache miss and TLB miss, and costs of boundary processing can be neglected.

However, traditional implementation and optimization methods of GEMM, as described above, cannot achieve optimal performance for small GEMM. Here are two reasons for this. First, the overhead of pack step in small GEMM is too high, as shown in Section \ref{sec:performance}. The advantages of pack step are no longer significant, but it results in high extra memory access overhead. Second, the costs of boundary processing are not neglected for small GEMM. Therefore, designing and implementing a method without pack steps and boundary processing is very necessary for achieving high performance of small GEMM.

This paper proposes an input-aware adaptive tuning framework(IAAT) for small GEMM to achieve near-optimal performance. IAAT has two stages, the install-time stage and the run-time stage. The install-time stage is responsible for auto-generating high-performance assembly kernels of different sizes. This stage automatically tunes kernels based on features of hardware to achieve optimal performance. The run-time stage's core is the input-aware adaptive tile algorithm, which tiles input matrices into some small blocks. This stage plays the role of runtime tuning by tiling matrix during program execution. Our performance evaluation shows that IAAT can achieve near-optimal performance when the size of input matrices is small as shown in Section \ref{sec:performance}.

Our contributions are summarized as follows:
\begin{itemize}
    \item We propose a template-based high-performance code auto-generation method to generate high-performance kernels for GEMM of different sizes in assembly language.
    \item We design an input-aware adaptive algorithm to divide input matrices into blocks in runtime to obtain a near-optimal solution. 
    \item We implement a high-performance input-aware adaptive tuning framework(IAAT) for small GEMM based on ARMv8.
\end{itemize}

The remainder of this paper is organized as follows. Section \ref{sec:relatedwork} presents related works. Section \ref{sec:framework} provides an overview of the framework. Section \ref{sec:installtime} and Section \ref{sec:runtime} introduces the details of two stages of IATT. Section \ref{sec:performance} presents the experimental results. Finally, Section \ref{sec:conclusions} concludes the paper.

\section{Related Works}
\label{sec:relatedwork}
Matrix multiplication has been optimized over years. Researchers utilize different methods and technologies to improve various matrix multiplication, such as tall and skinny matrix multiplication\cite{chen2019tsm2, rivera2021tsm2x, li2021autotsmm}, batches of matrix multiplication\cite{abdelfattah2020matrix,abdelfattah2019fast}, parallel matrix multiplication\cite{kang2020hpmax} and so on. 
For example,  tall and skinny matrix multiplication kernels are optimized by a flexible, configurable mapping scheme and outperform on an NVIDIA Volta GPGPU\cite{ernst2021performance}.
Lionel Eyraud-Dubois\cite{eyraud2018using} uses more general allocations to perform matrix multiplication on a heterogeneous node based on task-based runtime systems. 

Small GEMM are becoming more and more important in recent years. The optimization of small GEMM is introduced by many libraries, like LIBXSMM\cite{libxsmm}, BLIS\cite{blislib, blis2019}. LIBXSMM uses a code generator that has a built-in architectural model to auto-generate code. And the code runs well without requiring an auto-tuning phase by utilizing just-in-time compilation. BLIS uses the method of optimizing skinny matrix to optimize the small matrix and works well\cite{blis2019}. 

However, current methods and implementations of small GEMM cannot achieve near-optimal performance on ARMv8 platform. LIBXSMM and BLIS only focused on x86 CPU. Besides, BLIS tile algorithm cannot improve the performance of small GEMM to optimal performance of small GEMM\cite{blis2019}. And BLIS only implemented the small GEMM for single-precision and double-precision but not single-precision complex and double-precision complex. Distinguish from LIBXSMM and BLIS, we optimize all types of small GEMM for ARMv8 platform.

\begin{figure}[htbp]
\centering
\includegraphics[width=0.45\textwidth]{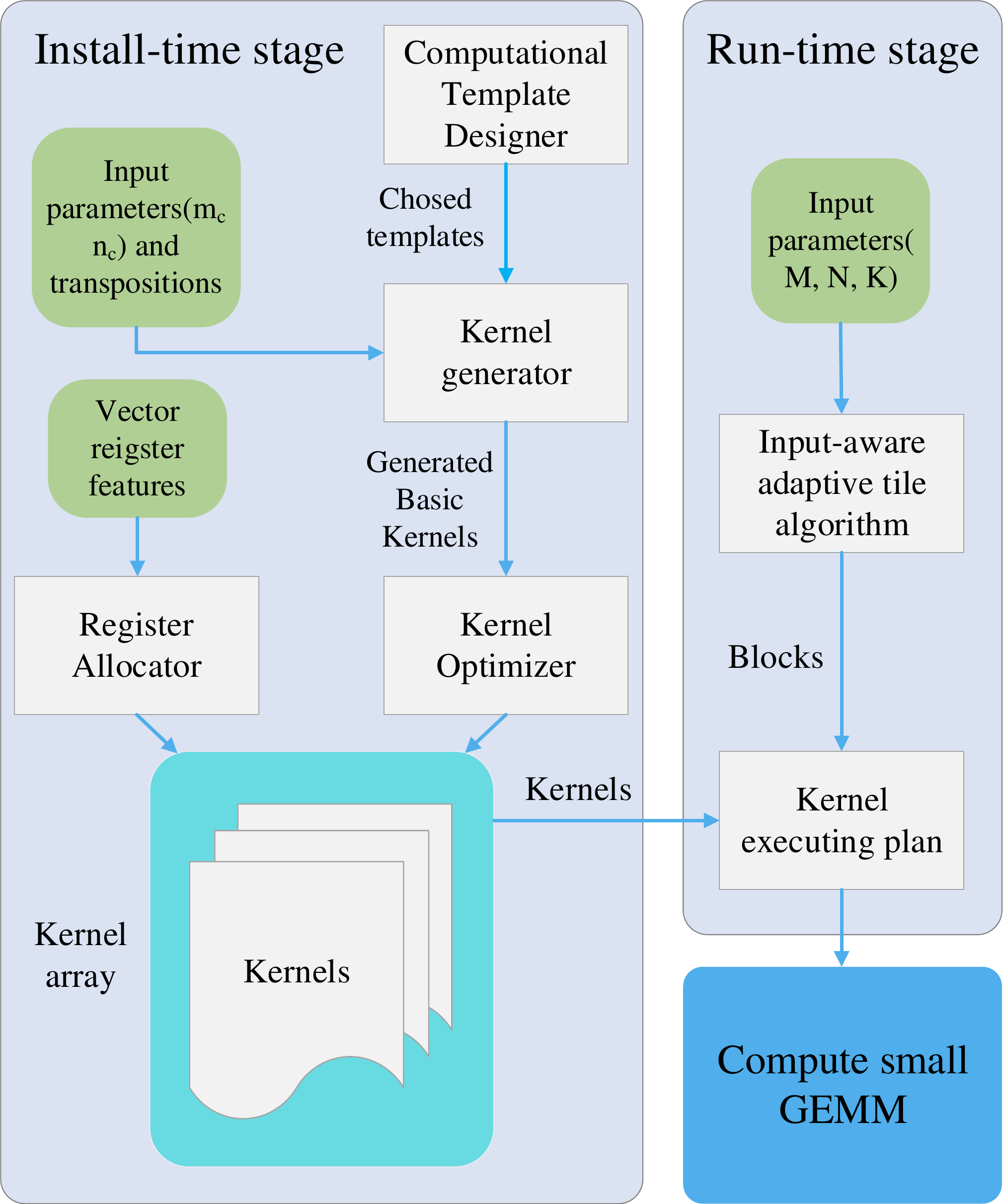}
\caption{The overall IAAT}
\label{fig:code framework}
\end{figure}

\section{Framework}
\label{sec:framework}
This section introduces the input-aware adaptive tuning framework(IAAT), as shown in Fig.\ref{fig:code framework}, with two stages, the install-time stage and the run-time stage to achieve near-optimal performance for small GEMM.

\subsection{The Install-Time Stage}
\label{subsec:framework_installtimestage}
The install-time stage auto-generates hundreds of kernels of different sizes. 
Pack step of the traditional method of GEMM makes data access continuous. So the traditional method of GEMM only needs one kernel to accomplish computation for different transpositions. After removing pack step, we have to use hundreds of kernels for different matrix sizes, types and transpositions. These kernels need lots of work to write by hand. Therefore, IAAT uses auto-generation to generate high-performance kernels in the install-time stage. This stage automatically tunes kernels based on features of the hardware to achieve optimal performance. The install-time stage utilizes four components to generate kernels:

\begin{itemize}
\item \textbf{Computational Template Designer} abstracts typical computing patterns of matrix multiplication as templates.
\item \textbf{Kernel Generator} designs a kernel generation algorithm, which utilizes templates from compute template designer to generate basic kernels of different sizes. 
\item \textbf{Register Allocator} allocates SIMD registers for kernels based on the size of kernel and SIMD register features.
\item \textbf{Kernel Optimizer} optimizes kernels from kernel generator to approach full potential power of the hardware.
\end{itemize}

\begin{table}[htbp]
\caption{ALL GENERATED KERNELS}
\footnotesize
\begin{center}
\setlength\tabcolsep{4pt}
\begin{tabular}{|c|c|c|c|c|}
\hline
 & \textbf{NN} & \textbf{NT} & \textbf{TN} & \textbf{TT} \\ \hline
\textbf{\rotatebox[origin=c]{270}{SGEMM}} & \begin{tabular}[c]{@{}l@{}}16$\times$\{1,2,3,4\}\\ 12$\times$\{1,2,...,6\}\\ 8$\times$\{1,2,...,8\}\\ 4$\times$\{1,2,...,13\}\\ 3$\times$\{1,2,...,13\}\\ 2$\times$\{1,2,...,13\}\\ 1$\times$\{1,2,...,13\}\end{tabular} & \begin{tabular}[c]{@{}l@{}}16$\times$\{1,2,3,4\}\\ 12$\times$\{1,2,...,8\}\\ 8$\times$\{1,2,...,8\}\\ 4$\times$\{1,2,...,20\}\\ 3$\times$\{1,2,...,24\}\\ 2$\times$\{1,2,...,28\}\\ 1$\times$\{1,2,...,32\}\end{tabular} & \begin{tabular}[c]{@{}l@{}}4$\times$\{1,2,3,4\}\\ 3$\times$\{1,2,3,4,5\}\\ 2$\times$\{1,2,...,7\}\\ 1$\times$\{1,2,...,10\}\end{tabular}                    & \begin{tabular}[c]{@{}l@{}}\{1,2,3,4\}$\times$16\\ \{1,2,...,6\}$\times$12\\ \{1,2,...,8\}$\times$8\\ \{1,2,...,13\}$\times$4\\ \{1,2,...,13\}$\times$3\\ \{1,2,...,13\}$\times$2\\ \{1,2,...,13\}$\times$1\end{tabular} \\ \hline
\textbf{\rotatebox[origin=c]{270}{DGEMM}} & \begin{tabular}[c]{@{}l@{}}8$\times$\{1,2,3,4\}\\ 4$\times$\{1,2,...,8\}\\ 3$\times$\{1,2,...,8\}\\ 2$\times$\{1,2,...,15\}\\ 1$\times$\{1,2,...,15\}\end{tabular}                                         & \begin{tabular}[c]{@{}l@{}}8$\times$\{1,2,3,4\}\\ 4$\times$\{1,2,...,8\}\\ 3$\times$\{1,2,...,8\}\\ 2$\times$\{1,2,...,20\}\\ 1$\times$\{1,2,...,20\}\end{tabular}                                         & \begin{tabular}[c]{@{}l@{}}4$\times$\{1,2,3,4\}\\ 3$\times$\{1,2,3,4,5\}\\ 2$\times$\{1,2,...,7\}\\ 1$\times$\{1,2,...,10\}\end{tabular}                    & \begin{tabular}[c]{@{}l@{}}\{1,2,3,4\}$\times$8\\ \{1,2,...,8\}$\times$4\\ \{1,2,...,8\}$\times$3\\ \{1,2,...,15\}$\times$2\\ \{1,2,...,15\}$\times$1\end{tabular}                     \\ \hline
\textbf{\rotatebox[origin=c]{270}{CGEMM}} & \begin{tabular}[c]{@{}l@{}}8$\times$\{1,2,3,4\}\\ 4$\times$\{1,2,...,9\}\\ 3$\times$\{1,2,...,9\}\\ 2$\times$\{1,2,...,12\}\\ 1$\times$\{1,2,...,20\}\end{tabular}                                         & \begin{tabular}[c]{@{}l@{}}8$\times$\{1,2,3,4\}\\ 4$\times$\{1,2,...,8\}\\ 3$\times$\{1,2,...,8\}\\ 2$\times$\{1,2,...,12\}\\ 1$\times$\{1,2,...,20\}\end{tabular}                                         & \begin{tabular}[c]{@{}l@{}}4$\times$\{1,2,...,9\}\\ 3$\times$\{1,2,...,9\}\\ 2$\times$\{1,2,...,12\}\\ 1$\times$\{1,2,...,20\}\end{tabular} & \begin{tabular}[c]{@{}l@{}}\{1,2,3,4\}$\times$8\\ \{1,2,...,9\}$\times$4\\ \{1,2,...,9\}$\times$3\\ \{1,2,...,12\}$\times$2\\ \{1,2,...,20\}$\times$1\end{tabular}                      \\ \hline
\textbf{\rotatebox[origin=c]{270}{ZGEMM}} & \begin{tabular}[c]{@{}l@{}}4$\times$\{1,2,3,4\}\\ 3$\times$\{1,2,3,4\}\\ 2$\times$\{1,2,...,7\}\\ 1$\times$\{1,2,...,10\}\end{tabular}                                                              & \begin{tabular}[c]{@{}l@{}}4$\times$\{1,2,3,4\}\\ 3$\times$\{1,2,3,4\}\\ 2$\times$\{1,2,...,7\}\\ 1$\times$\{1,2,...,10\}\end{tabular}                                                              & \begin{tabular}[c]{@{}l@{}}4$\times$\{1,2,3,4\}\\ 3$\times$\{1,2,3,4\}\\ 2$\times$\{1,2,...,7\}\\ 1$\times$\{1,2,...,10\}\end{tabular}                      & \begin{tabular}[c]{@{}l@{}}\{1,2,3,4\}$\times$4\\ \{1,2,3,4\}$\times$3\\ \{1,2,...,7\}$\times$2\\ \{1,2,...,10\}$\times$1\end{tabular}                                           \\ \hline
\end{tabular}
\end{center}
\footnotesize
We define the kernel for different matrix types and different transpositions. The SGEMM/DGEMM/CGEMM/ZGEMM represent single-precision matrix multiplication, double-precision matrix multiplication, single-precision complex matrix multiplication, double-precision complex matrix multiplication. Each type has four transpositions, NN, NT, TN, and TT. For example, NT means matrix A is not transposed and matrix B is transposed. We will also use abbreviations like SGEMM\_TN, which means input matrix type is single and matrix A is transposed and matrix B isn't transposed. We acquiescence the matrix in column-major order.
\label{tab:ALL GENERATED KERNELS}
\end{table}

TABLE \ref{tab:ALL GENERATED KERNELS} shows all kernels we defined in this paper, which are completely auto-generated. All these kernels construct basic computation of small GEMM and form a kernel array, which is directly invoked by the run-time stage.

\subsection{The Run-Time Stage}
The run-time stage tiles input matrices A, B, and C into blocks and generates a near-optimal small GEMM kernel executing plan.
The costs of boundary processing can be neglected for GEMM. However, for small GEMM, the costs of boundary processing are high and cannot be neglected. To reduce or eliminate boundary processing, an algorithm is required, which can tile input matrices into optimal blocks with less boundary processing. The core of the run-time stage is the input-aware adaptive tile algorithm. This algorithm tiles input matrices into optimal blocks according to the size of kernels from the install-time stage. These blocks are tuned according to input matrix sizes, types and transpositions. Therefore, the run-time stage plays the role of runtime tuning. Then this stage connects the kernel to form a sequence of kernels, which is called the kernel executing plan. Finally, IAAT computes the small GEMM based on this kernel executing plan.
 
\section{The Install-Time Stage}
\label{sec:installtime}
This section focuses on the install-time stage, which auto-generates hundreds of kernels of different sizes. Below we introduce four components of the install-time stage as shown in Fig.\ref{fig:code framework}.

\subsection{Computational Template Designer}
To construct main calculation of GEMM kernel, we introduce computational template designer. The computational template designer extracts typical computing patterns of matrix multiplication as templates, which are shown in TABLE \ref{tab:The Kernel Computational Templates}.
\begin{itemize}
\item \textbf{sfmlas} and \textbf{dfmlas} represent a vector-scalar multiply-add operation.
\item \textbf{sfmlav} and \textbf{dfmlav} represent a vector-vector multiply-add operation.
\item \textbf{sfmlss} and \textbf{dfmlss} represent a vector-scalar multiplication and subtraction.
\item \textbf{sfnegv} and \textbf{dfnegv} are used to invert values in register.
\item \textbf{sfcmlas} and \textbf{dfcmlas} represent a vector-scalar complex multiply-add operation.
\item \textbf{sfcmlav} and \textbf{dfcmlav} represent a vector-vector complex multiply-add operation.
\end{itemize}

\begin{table}[htbp]
\caption{Kernel Computational Templates}
\scriptsize
\begin{center}
\begin{tabular}{|l|l|}
\hline
\begin{tabular}[c]{@{}l@{}}\textbf{sfmlas}(out, in1, in2, index):\\\  
fmla out.4s, in1.4s, in2.s{[}index{]} 
\end{tabular}                        & 
\begin{tabular}[c]{@{}l@{}}\textbf{dfmlas}(out, in1, in2, index):\\\  
fmla out.2d, in1.2d,  in2.d{[}index{]}
\end{tabular} \\ \hline
\begin{tabular}[c]{@{}l@{}}\textbf{sfmlav}(out, in1, in2):\\\  
fmla out.4s, in1.4s, in2.4s
\end{tabular}                       & 
\begin{tabular}[c]{@{}l@{}}\textbf{dfmlav}(out, in1, in2):\\\  
fmls out.2d, in1.2d,  in2.2d
\end{tabular} \\ \hline
\begin{tabular}[c]{@{}l@{}}\textbf{sfmlss}(out, in1, in2, index):\\\  
fmls out.4s, in1.4s, in2.s{[}index{]}
\end{tabular}                       & 
\begin{tabular}[c]{@{}l@{}}\textbf{dfmlss}(out, in1, in2, index):\\\  
fmla out.2d, in1.2d, in2.d{[}index{]}
\end{tabular} \\ \hline
\begin{tabular}[c]{@{}l@{}}\textbf{sfnegv}(out, in1):\\\ 
fneg out.4s, in1.4s
\end{tabular}                       & 
\begin{tabular}[c]{@{}l@{}}\textbf{dfnegv}(out, in1):\\\  
fneg out.2d, in1.2d
\end{tabular} \\ \hline
\begin{tabular}[c]{@{}l@{}}\textbf{sfcmlas}(out, in1, in2, index, rot[2]):\\\  
fcmla out.4s, in1.4s,  in2.s{[}index{]}, rot[0] \\\ 
fcmla out.4s, in1.4s,  in2.s{[}index{]}, rot[1]
\end{tabular}                       & 
\begin{tabular}[c]{@{}l@{}}\textbf{dfcmlas}(out, in1, in2, rot[2]):\\\  
fcmla out.2d, in1.2d,  in2.2d, rot[0] \\\ 
fcmla out.2d, in1.2d,  in2.2d, rot[1]
\end{tabular} \\ \hline
\begin{tabular}[c]{@{}l@{}}\textbf{sfcmlav}(out, in1, in2, rot[2]):\\\  
fcmla out.4s, in1.4s,  in2.4s, rot[0] \\\ 
fcmla out.4s, in1.4s,  in2.4s, rot[1]
\end{tabular}                       & 
\begin{tabular}[c]{@{}l@{}}\textbf{dfcmlav}(out, in1, in2, rot[2]):\\\  
fcmla out.2d, in1.2d,  in2.2d, rot[0] \\\ 
fcmla out.2d, in1.2d,  in2.2d, rot[1]
\end{tabular} \\ \hline
\end{tabular}
\label{tab:The Kernel Computational Templates}
\end{center}
\end{table}

\subsection{Kernel Generator}
\label{subsec:kernel generator}
Kernel generator is responsible for generating kernels. These kernels are used to compute $C_c=A_c\times B_c+Cc$. Here $A_c$, $B_c$, and $C_c$ are blocks of input matrices A, B, and C. And they are $m_c \times k_c$,  $k_c \times n_c$, and $m_c \times n_c$ matrices, respectively. The algorithm of kernel generator takes size of $C_c$ as input and outputs high-performance kernel in assembly language.

Kernel generator generates two kinds of subkernels for ping-pang operation. The ping-pang operation is an optimization method that split the multiplication into two stages, M1 and M2 stages. There are two types of ping-pang operations. In the first type, each stage of ping-pang operation multiplies a column of block $A_c$ and a row of block $B_c$ and loads the next column of block $A_c$ and next row of block $B_c$. In the second type, each stage multiplies a column of block $A_c$ and a row of block $B_c$, M1 stage loads the next column of block $A_c$ and two rows of block $B_c$, and M2 stage loads the next column of block $A_c$. And the performance difference between these two types is not too much.

The kernel generator algorithms for various input matrix types and transpositions are similar. We only discuss SGEMM\_NN kernel generator shown in Algorithm \ref{alg:kernel generation of SGEMMNN}.

SGEMM\_NN kernel generator generates two subkernels in lines 6-12 and 14-19. The first subkernel loads a column of $A_c$ and two rows of $B_c$ in lines 6-7 and the second subkernel loads a column of $A_c$ in line 14. Each subkernel multiplies a column of $A_c$ and a row of $B_c$ by utilizing \textbf{sfmlas} in lines 8-12 and 15-19.

After two kinds of subkernels of SGEMM\_NN are generated, the kernel generator invokes these two subkernels in a loop on the $k_c$ dimension and completes the generation of SGEMM\_NN kernel.

\begin{algorithm}[h]
\caption{kernel generator of SGEMM\_NN}
\label{alg:kernel generation of SGEMMNN}
\begin{algorithmic}[1]
	    \REQUIRE $m_c$, $n_c$: the size of the input kernel
	    \ENSURE $kernel$
	    \STATE $Cregs \gets \{C_1, C_2,..., C_{m_{\left \lceil m_c/4 \right \rceil}n_c}\}$
	    \STATE $A1regs \gets \{A_1, A_2,..., A_{\left \lceil m_c/4 \right \rceil}\}$
	    \STATE $A2regs \gets \{A_{\left \lceil m_c/4 \right \rceil+1},$ $A_{\left \lceil m_c/4 \right \rceil+2},$ $...,$ $A_{2\left \lceil m_c/4 \right \rceil}\}$
	    \STATE $Bregs \gets \{B_1, B_2,..., B_{n_c}\}$
	    \STATE //first subkernel
	    \STATE load next column of block $A_c$ to $A2regs$
	    \STATE load two rows of block $B_c$ to $Bregs$
	    \FOR{$i$ $\gets$ $0$ $to$ $n_c$}
	        \FOR{$j$ $\gets$ $0$ $to$ $\left \lceil m_c/4 \right \rceil$}
	            \STATE \textbf{sfmlas}($Cregs{[}i\left \lceil m_c/4 \right \rceil+j{]}$, $A1regs{[}j{]}$, $Bregs{[}i{]}$, 0)
	        \ENDFOR
	    \ENDFOR
	    \STATE //second subkernel
	    \STATE load next column of block $A_c$ to $A1regs$
	    \FOR{$i$ $\gets$ $0$ $to$ $n_c$}
	        \FOR{$j$ $\gets$ $0$ $to$ $\left \lceil m_c/4 \right \rceil$}
	            \STATE \textbf{sfmlas}($Cregs{[}i\left \lceil m_c/4 \right \rceil+j{]}$, $A2regs{[}j{]}$, $Bregs{[}i{]}$, 1)
	        \ENDFOR
	    \ENDFOR
\end{algorithmic}
\end{algorithm}

\subsection{Register Allocator}
\label{subsec:register allocator}
The allocation of registers is very important for the performance of small GEMM. Hence, we need to define the strategies of register allocation for different kernels. The work of our paper is mainly carried out on ARMv8 platform, which contains 32 128-bit SIMD registers.

The basic idea behind the register allocator is to divide all registers into three groups. $A_c$ register group contains two columns of $A_c$; $B_c$ register group contains two rows of $B_c$ for ping-pang operation; $C_c$ register group holds the whole block $C_c$.

Allocation of the $A_c$ register group has four main strategies, ANTwoCC, ATEachCTwo, ATEachCOne, and ATTwoRR. 
\begin{itemize}
    \item \textbf{ANTwoCC} is for loading two columns of $A_c$ to registers. It allocates $2\left \lceil m_c/elenum \right \rceil$ registers, the $elenum$ means the number of elements that a register can store.
    \item \textbf{ATEachCTwo} is for loading first two data of each column of transposed $A_c$ to two registers. It allocates a total of $2m_c$ registers.
    \item \textbf{ATEachCOne} is for loading first two data of each column of transposed $A_c$ to one register. It requires a total of $m_c$ registers for single-precision, double-precision, and single-precision complex. As for double-precision complex, it requires a total of $2m_c$ registers. 
    \item \textbf{ATTwoRR} is for loading two rows of transposed $A_c$ to registers. It allocates $2\left \lceil m_c/elenum \right \rceil$ registers.
\end{itemize}

The strategies of allocating $B_c$ register group are BTTwoCC, BNEachCTwo, BNEachCOne, and BNTwoRR corresponding to ANTwoCC, ATEachCTwo, ATEachCOne, and ATTwoRR. This is because load methods of $A_c$ are the same as load methods of $B_c$.

The strategy of allocating the C register group is allocating $\left \lceil m_c\times n_c/elenum \right \rceil$ registers.

The register allocator has one special strategy for TN transposition that allocates $2m_c$ registers for $A_c$ and $2n_c$ registers for $B_c$. This transposition makes memory access to $A_c$ and $B_c$ discontinuous. So we cannot vectorize small GEMM for this transposition. Therefore, the methods of loading data are load data from each column of $A_c$ by columns and load data from each column of $B_c$ by columns.

\subsection{Kernel Optimizer}
After kernels are generated, kernels will be optimized as follows.

\paragraph{Instruction Choice} Computational template designer utilizes the FMA instruction instead of $mul$ or $add$ because there usually are fused multiply-add(FMA) units in hardware. Besides, we prioritize the $ldp$ and $ldr$ instructions because these two instructions are relatively high-performance.

\paragraph{Instruction Order} The loading instructions are interspersed among the computing instructions. It makes better use of the instruction pipeline to avoid pipeline stalling.

\paragraph{Ping-Pang Operation} As described in Subsection \ref{subsec:kernel generator}, this optimization utilizes computing instruction to hide the delay of loading instructions.

\section{The Run-Time Stage}
\label{sec:runtime}
This section introduces the run-time stage. This stage first tile input matrices and then construct a kernel executing plan to compute small GEMM. The input-aware adaptive tile algorithm is the core of this stage.

\subsection{Input-Aware Adaptive Tile Algorithm}
The input-aware adaptive tile algorithm first tiles input matrix C into some small blocks. Each block has the same size as one of the generated kernels. Then, this algorithm tiles matrices A and B based on tiled blocks of C. This algorithm is based on the three principles listed below.

\begin{algorithm}[h]
	\caption{SGEMM\_NN Tile Algorithm}
	\label{alg:SGEMMNNtileAlg}
	\footnotesize
	\begin{algorithmic}[1]
	    \REQUIRE $M,N,K$: the sizes of input matrices, $kernels$: array of all sorted SGEMM\_NN kernels from TABLE \ref{tab:ALL GENERATED KERNELS}
	    \ENSURE $blocksC$[], $blocksA$[], $blocksB$[]
	    \IF {$N \le 13$}
	        \STATE $m[] \gets (m_1, I)$, $m_1$ is the largest $m_c$ of kernel that's $n_c$ is equal $N$ and $I$ is an integer and make sure the $m_1I \le M$
	        \STATE $n[] \gets [(N, 1)]$
            \IF{$m_1I < M$}
                \STATE $m$.append($(M-m_1I, 1)$)
                \STATE $n$.append($[(N, 1)]$)
            \ENDIF
        \ELSE
            \IF{$M < 8$}
                \STATE $m[] \gets TileSingleDim(M, [1,2,3,4])$
                \STATE $n[] \gets [TileSingleDim(N, [1,2,...,13])]$
                \IF{size of $m$ == 2}
                    \STATE $n.append([TileSingleDim(N, [1,2,...,13])])$
                \ENDIF
            \ELSIF{$M == 9$}
                \STATE $m[] \gets (4, 1), (3, 1), (2, 1)$
                \STATE $n[] \gets [TileSingleDim(N, [1,2,...,13])],$ $[TileSingleDim$ $(N,$ $[1,2,...,13])],$ $[TileSingleDim(N,$ $ [1,2,...,13])]$
            \ELSIF{$M < 12$}
                \STATE $m[] \gets (8, 1), (M - 8, 1)$
                \STATE $n[] \gets [TileSingleDim(N,[1,2,...,8])],$ $[TileSingleDim$ $(N,$ $[1,2,...,13])]$
            \ELSIF{$M == 12$}
                \STATE $m[] \gets (12, 1)$
                \STATE $n[] \gets [TileSingleDim(N, [1,2,3,4,5,6])]$
            \ELSE
                \STATE $m_1[] \gets (4, \lfloor M/4 \rfloor)$
                \STATE $m_2[] \gets (M-4\lfloor M/4 \rfloor, 1)$
                \STATE $n_2[] \gets [TileSingleDim(N, [1,2,...,13])]$
                \IF{$M-4\lfloor M/4 \rfloor == 1$}
                    \STATE $m_1[] \gets (4, \lfloor M/4 \rfloor-1)$
                    \STATE $m_2[] \gets (3, 1), (2, 1)$
                    \STATE $n_2[] \gets$ $[TileSingleDim(N,$ $[1,2,...,8])],$ $[TileSingleDim$ $(N,$ $[1,2,...,13])]$
                \ENDIF
                \STATE $m8\gets ExtendTo8(m_1)$
                \STATE $m16\gets ExtendTo16(m_1)$
                \STATE $n8[]$ and $n16[]$ $\gets$ tile $N$ by $m8[]$ and $m16[]$
                \STATE $blocksC1\gets Combine(m8, n8)$
                \STATE $blocksC2\gets Combine(m16, n16)$
                \STATE $blocksC\gets CompareLessMemops(blocksC1, blocksC2)$
                \STATE $blocksC$.append(combine($m_2$, $n_2$))
                \STATE return
            \ENDIF
	    \ENDIF
	    \STATE $blocksC \gets Combine(m, n)$
	    \STATE $blocksA[] \gets$ tile matrix A according to $blocksC[]$
	    \STATE $blocksB[] \gets$ tile matrix B according to $blocksC[]$
	\end{algorithmic}
\end{algorithm}

\paragraph{Bigger Block Size} Smaller blocks cause matrices A and B to be repeatedly loaded more times. The larger the block size, the lower the number of repetitions.

\paragraph{Minimal Memops} Different tiling methods have the same amount of computing instructions but different numbers of loading instructions. Therefore, the optimal tiling method is tiling matrices into blocks with the fewest loading instructions. The tiled blocks for matrix C are supposed to be $m_0 \times n_0$, $m_1 \times n_1$, ..., $m_i \times n_i$. The $m_i \times n_i$ is size of tiled block. This tiling method have a total of $(m_0+n_0+m_1+n_1...+m_a+n_a)K+2mn$ data to access from L2 cache to register. So the value of $(m_0+n_0+m_1+n_1...+m_a+n_a)$ should be preserved to a bare minimum.

\paragraph{SIMD Friendly} The dimension of block, that data is continuous, can be divisible by the length of SIMD register.

The pseudo-code of SGEMM\_NN tile algorithm are shown in Algorithm \ref{alg:SGEMMNNtileAlg}. The outline of this algorithm is below:

When $N \le 13$, we let $n_c=N$ and make $m_c$ the maximum value that $m_c$ can be taken in lines 1-7. When $N > 13$, we first tile $M$ into multiples of 4 and use 1, 2, 3 to supplement the deficiency, and
then tile $N$ into maximum value that $n_c$ can be taken according to the result of $M$'s tile in lines 9-42. Besides, when $M > 12$, $M$ can be tiled by 8 or 16 and we compare which one is better by counting the number of loading instructions and choose that. Then, we combine the two tiled dimensions $m[]$ and $n[]$ into $blocksC$. Finally, we tile matrices A and B into $blocksA$ and $blocksB$ according to blocks of matrix C.

$TileSingleDim$ algorithm, as shown in line 10, is for tiling a single dimension. It takes two input parameters: the length that you want to tile, and the array of lengths that you used to tile. This algorithm outputs array $(dim, nums)$ means $dim$ is repeated $nums$ times. We tile $dim$ into $nums_1I+nums_2...+nums_i$ and the bigger $nums_1$, the better. And if $nums_i$ is too small, this algorithm will average $nums_{i-1}$ and $nums_i$.

For various types and transpositions, the specific tile algorithm is changed slightly. But the basic ideas are consistent as shown above.

\begin{figure}[htbp]
\centering
\subfigure[traditional tiling method]{
\includegraphics{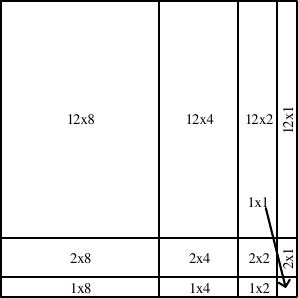}
\label{fig:fig1}
}
\quad
\subfigure[new tiling method]{
\includegraphics{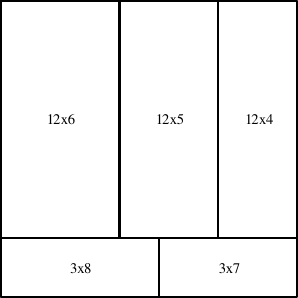}
\label{fig:fig2}
}
\caption{Schematic sketch of tiling method for a $15 \times 15$ SGEMMNN matrix}
\end{figure}

For SGEMM\_NN $15\times 15\times K$ matrix, the traditional tiling method is showed in Figure \ref{fig:fig1}. This method needs to load $105k+450$ data from L2 cache to register. And, our method tile SGEMM\_NN is showed in Figure \ref{fig:fig2}. This tiling method needs to load $72K+450$ data. The amount of data loaded by the traditional method is 45\% more than that of our method.

\subsection{Kernel Executing Plan}
After input matrices are tiled, IAAT constructs a kernel executing plan by connecting kernels, which correspond to the sizes of tiled blocks. Finally, IAAT executes this plan to compute small GEMM.

\section{Performance Evaluation}
\label{sec:performance}
In this section, we analyze small GEMM’s performance on ARMv8 platform as listed in TABLE \ref{tab:Experimental Environment Of ARMv8 platform}. We compared IAAT with currently state-of-the-art BLAS libraries: OpenBLAS, ARMPL, and BLIS. GEMM in these libraries is well optimized. Our work supports four data types: single-precision, double-precision, single-precision complex, and double-precision complex. Each data type supports four transpositions: NN, NT, TN, TT. Thus, we compared 16 kinds of small GEMMs. We use Equation \ref{equ:sd} to evaluate performance of SGEMM and DGEMM and Equation \ref{equ:cz} to evaluate performance of CGEMM and ZGEMM.

\begin{table}[htbp]
\centering
\caption{Experimental Environment Of ARMv8 platform}
\renewcommand{\arraystretch}{1.2}
\setlength{\tabcolsep}{10pt}
\begin{tabular}{lcc}
\hline
\multicolumn{1}{l}{Hardware} & CPU      & Kunpeng920 \\
\hline
&Arch.    & ARMv8.2    \\
&Freq.    & 2.6GHz     \\
&SIMD     & 128bits    \\
&L1 cache & 4MiB       \\
&L2 cache & 32MiB      \\
\hline
\multicolumn{1}{l}{Software} &Compiler & GCC7.5     \\
&OpenBLAS & 0.3.13    \\
&ARMPL    & 21.0       \\
&BLIS     & 0.81       \\
\hline
\end{tabular}
\label{tab:Experimental Environment Of ARMv8 platform}
\end{table}

\begin{eqnarray}
GFLOPS&=&\frac{2\times M \times N\times K}{t} \label{equ:sd}\\
GFLOPS&=&\frac{2\times 4 \times M \times N\times K}{t} \label{equ:cz}
\end{eqnarray}

\begin{figure}[htbp]
\centering
\includegraphics[width=3.4in]{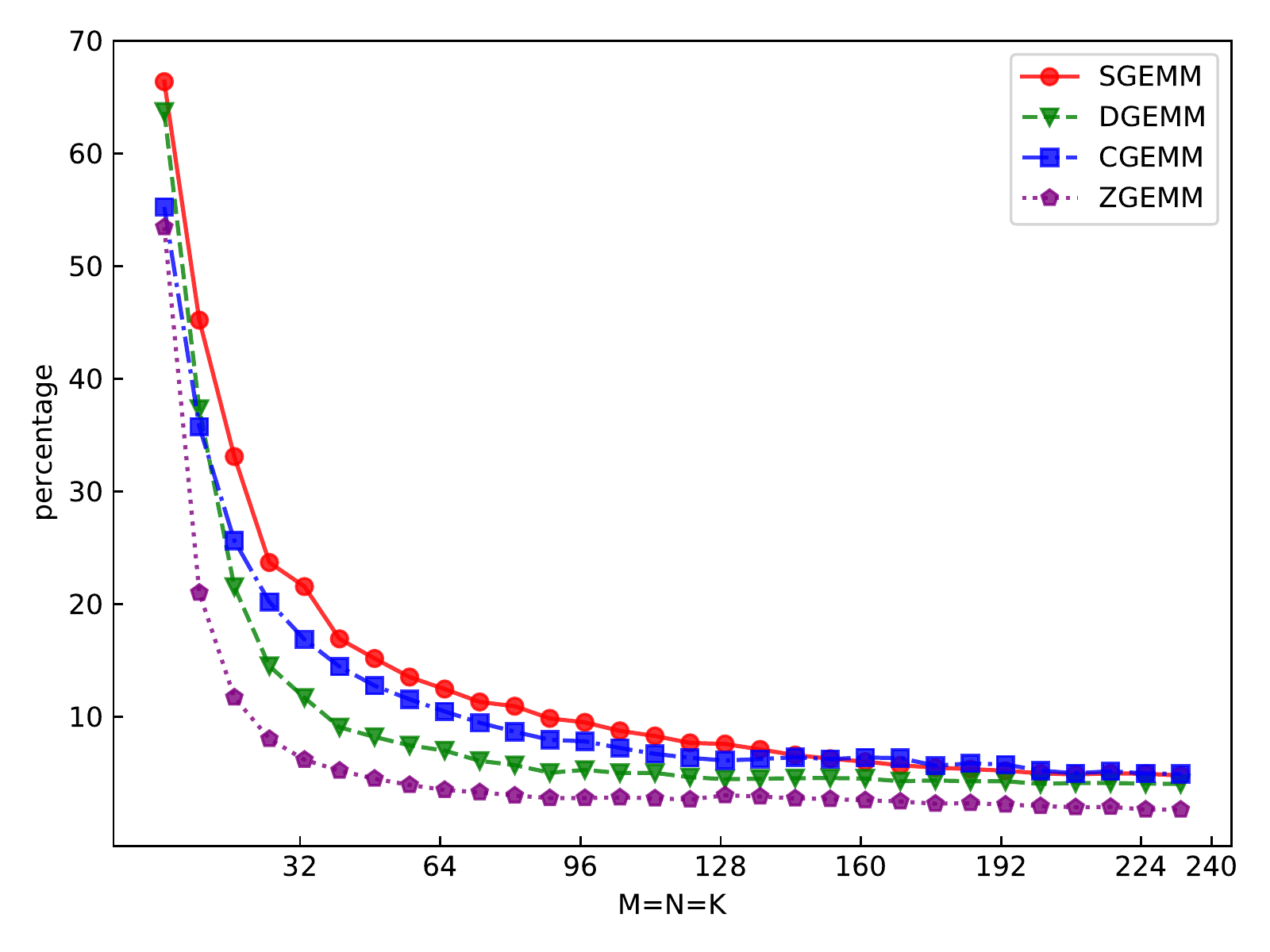}
\caption{Pack step cost proportion}
\label{fig:pack}
\end{figure}

Fig.\ref{fig:pack} shows proportion of pack step cost in traditional implementation of GEMM. It shows that the proportion of pack step cost can reach 67\% when input matrices are very small. As the size of input matrices increases, the proportion decreases exponentially. When input matrices are large enough, the proportion is near 3\%.

\begin{figure*}[htbp]
\centering
\subfigure[NN]{
\begin{minipage}[t]{0.23\linewidth}
\centering
\includegraphics[width=4.1cm]{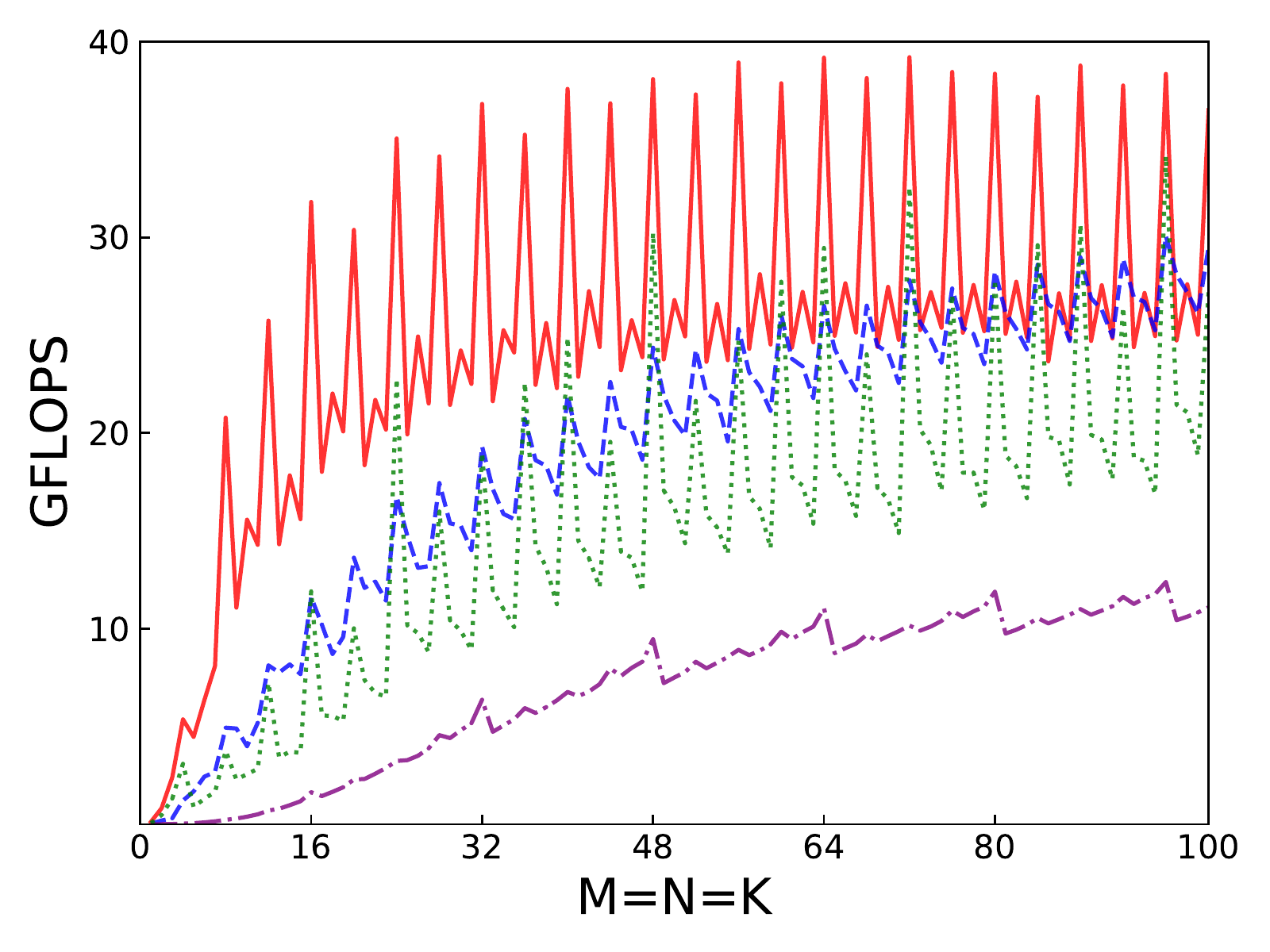}
\end{minipage}
\label{fig:snn}
}
\hspace{-0.5cm}
\subfigure[NT]{
\begin{minipage}[t]{0.23\linewidth}
\centering
\includegraphics[width=4.1cm]{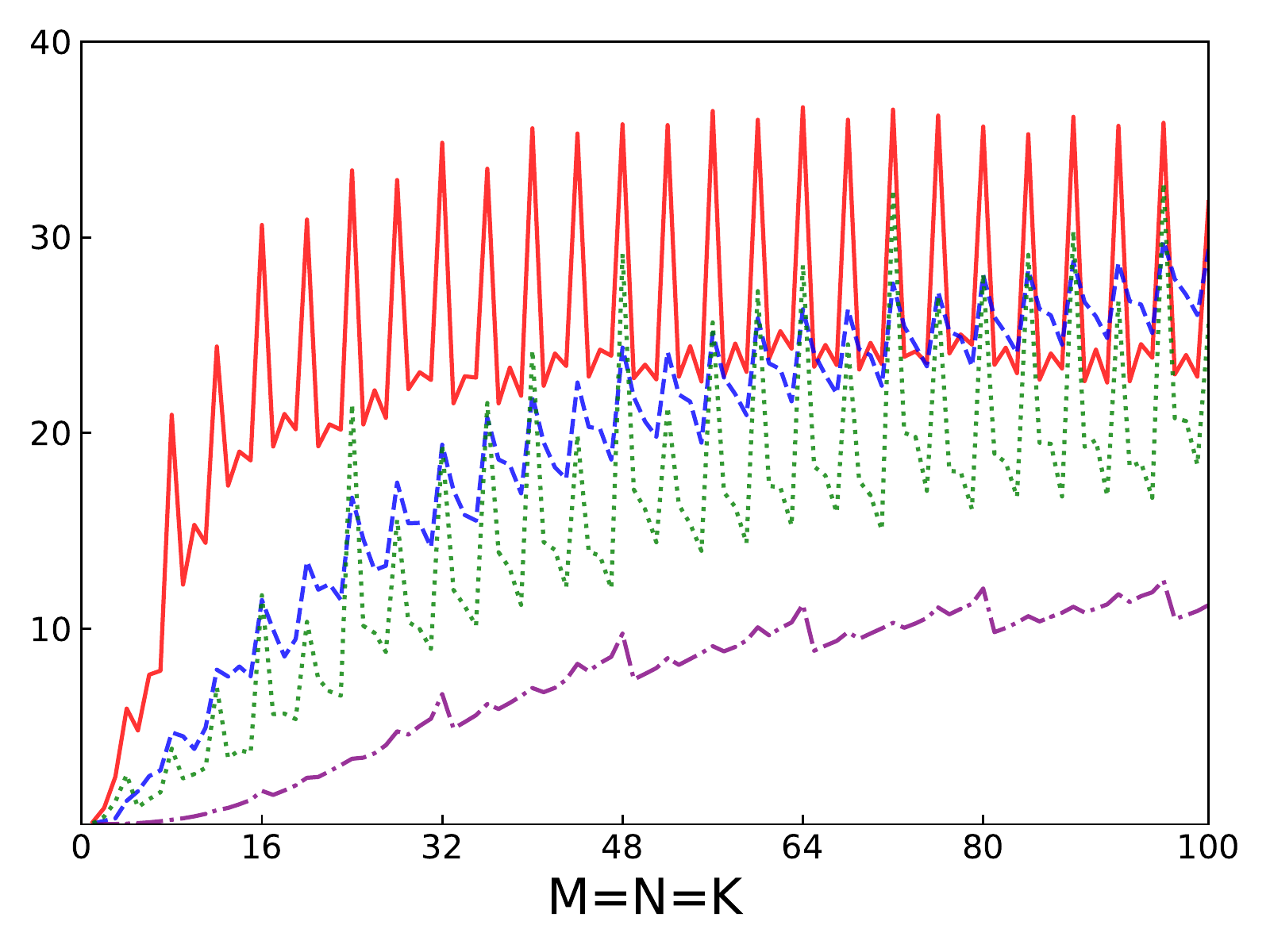}
\end{minipage}
\label{fig:snt}
}
\hspace{-0.5cm}
\subfigure[TN]{
\begin{minipage}[t]{0.23\linewidth}
\centering
\includegraphics[width=4.1cm]{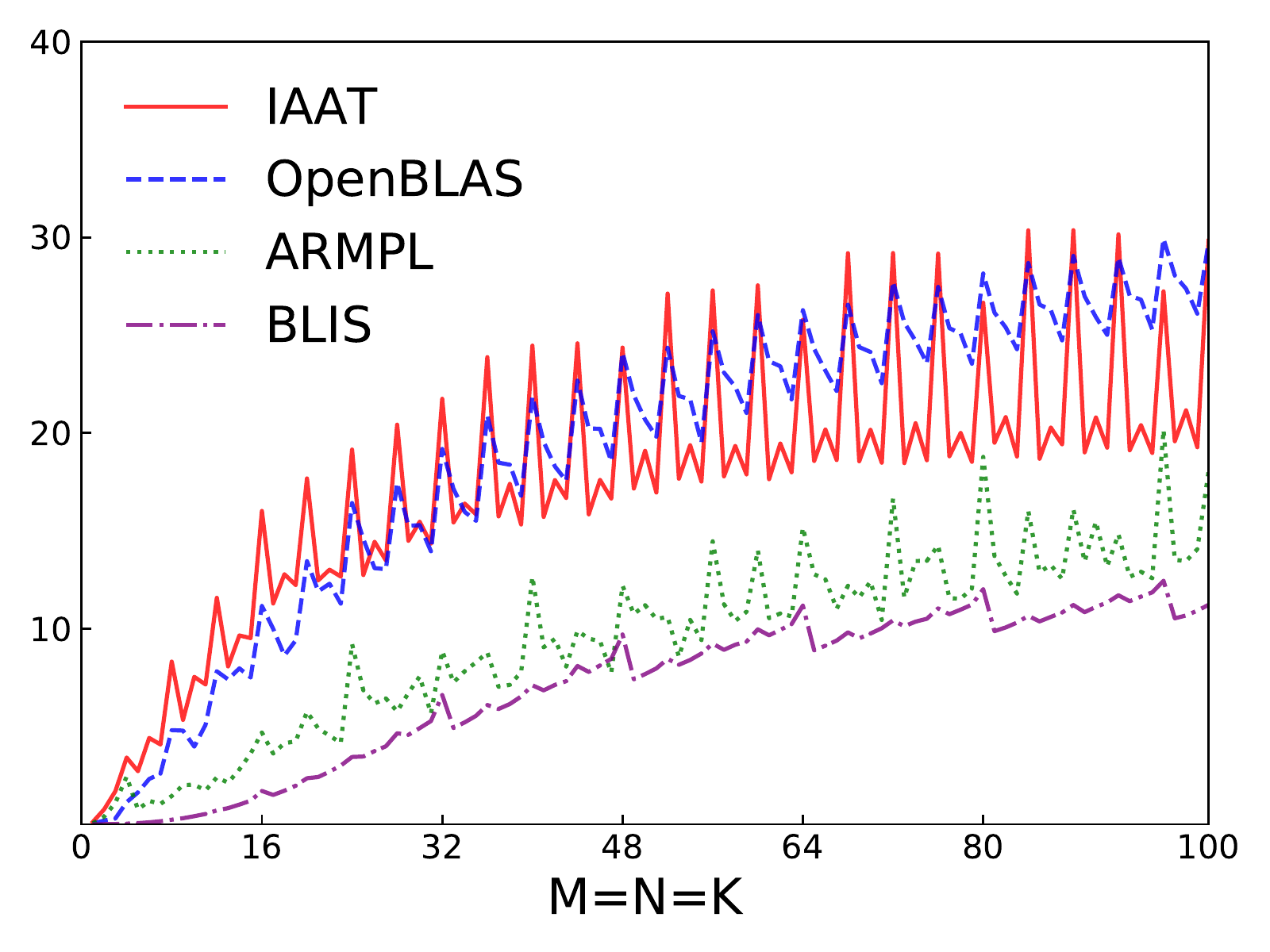}
\end{minipage}
\label{fig:stn}
}
\hspace{-0.5cm}
\subfigure[TT]{
\begin{minipage}[t]{0.23\linewidth}
\centering
\includegraphics[width=4.1cm]{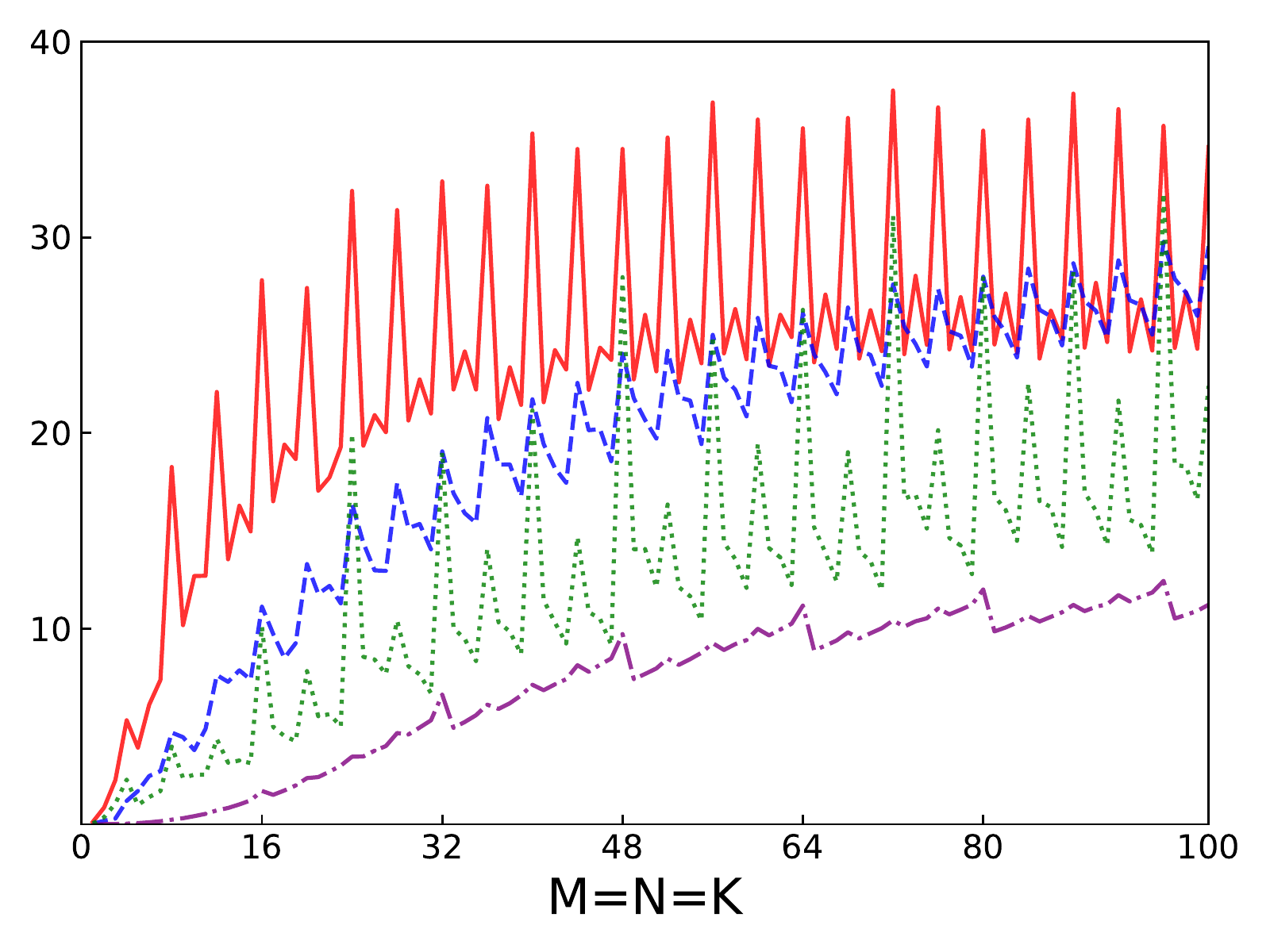}
\end{minipage}
\label{fig:stt}
}
\centering
\caption{Performance evaluation of IAAT vs. OpenBLAS, BLIS, ARMPL for SGEMM}
\label{fig:sgemm}
\end{figure*}

\begin{figure*}[htbp]
\centering
\subfigure[NN]{
\begin{minipage}[t]{0.23\linewidth}
\centering
\includegraphics[width=4.1cm]{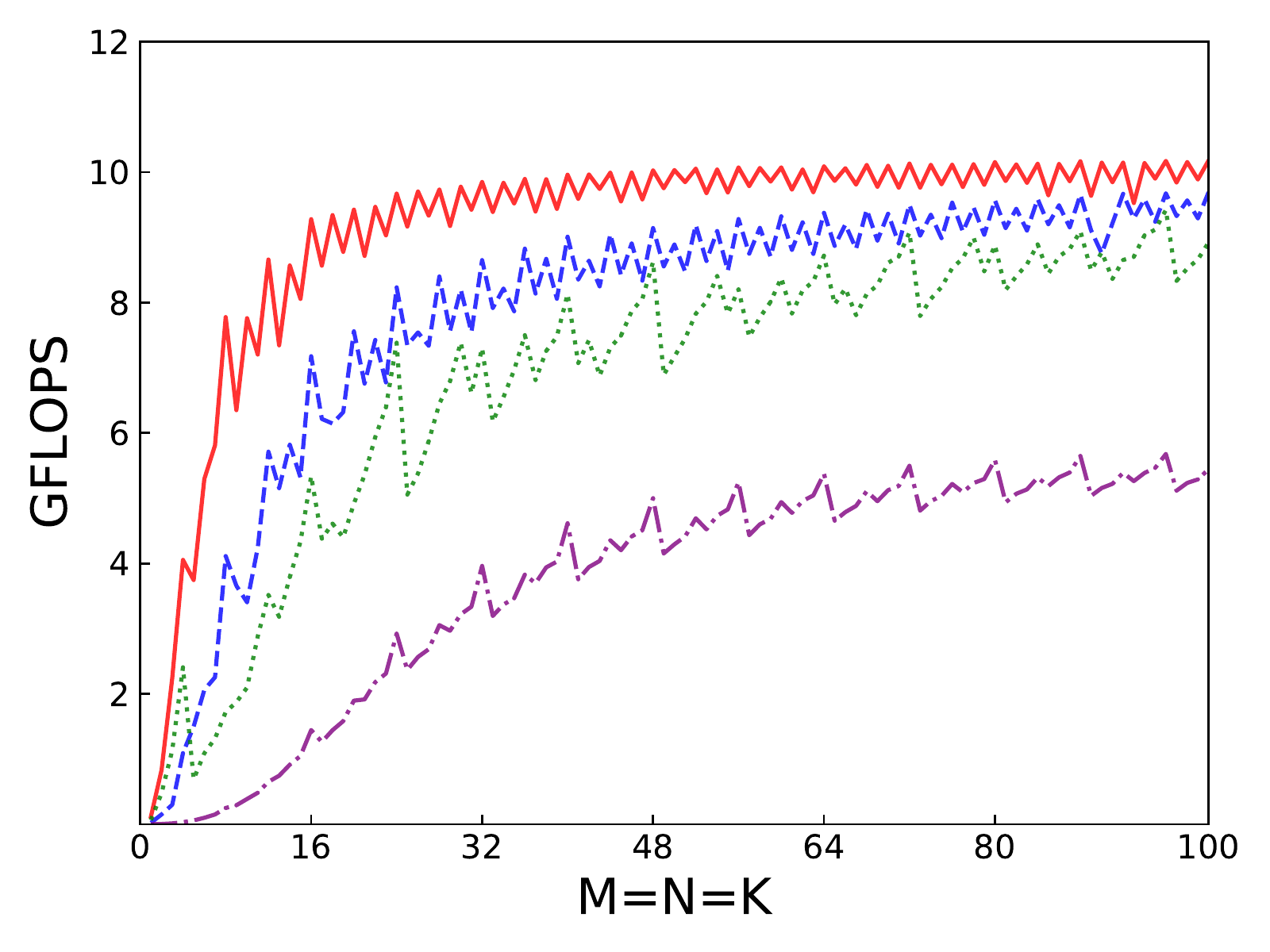}
\end{minipage}
\label{fig:dnn}
}
\hspace{-0.5cm}
\subfigure[NT]{
\begin{minipage}[t]{0.23\linewidth}
\centering
\includegraphics[width=4.1cm]{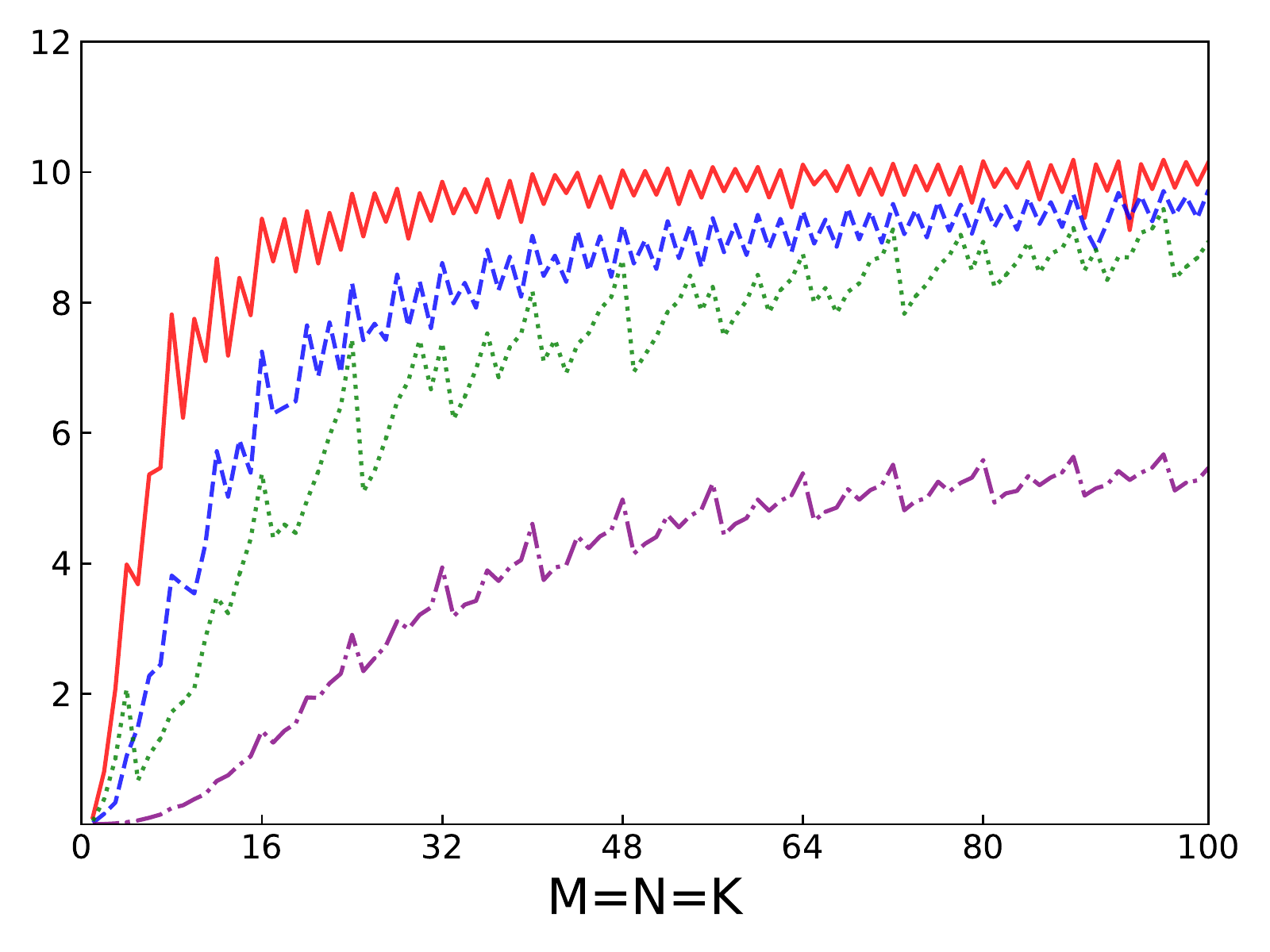}
\end{minipage}
\label{fig:dnt}
}
\hspace{-0.5cm}
\subfigure[TN]{
\begin{minipage}[t]{0.23\linewidth}
\centering
\includegraphics[width=4.1cm]{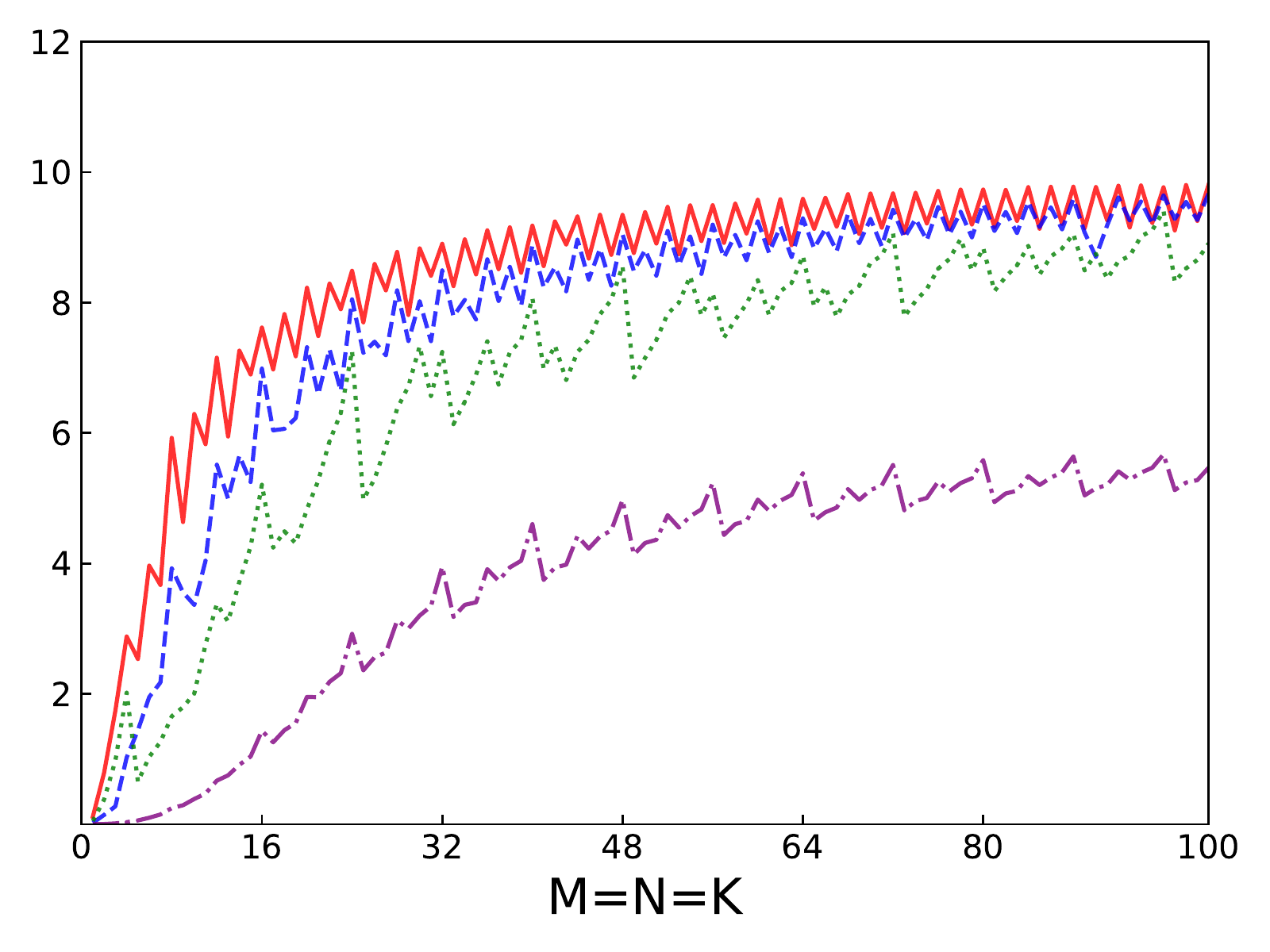}
\end{minipage}
\label{fig:dtn}
}
\hspace{-0.5cm}
\subfigure[TT]{
\begin{minipage}[t]{0.23\linewidth}
\centering
\includegraphics[width=4.1cm]{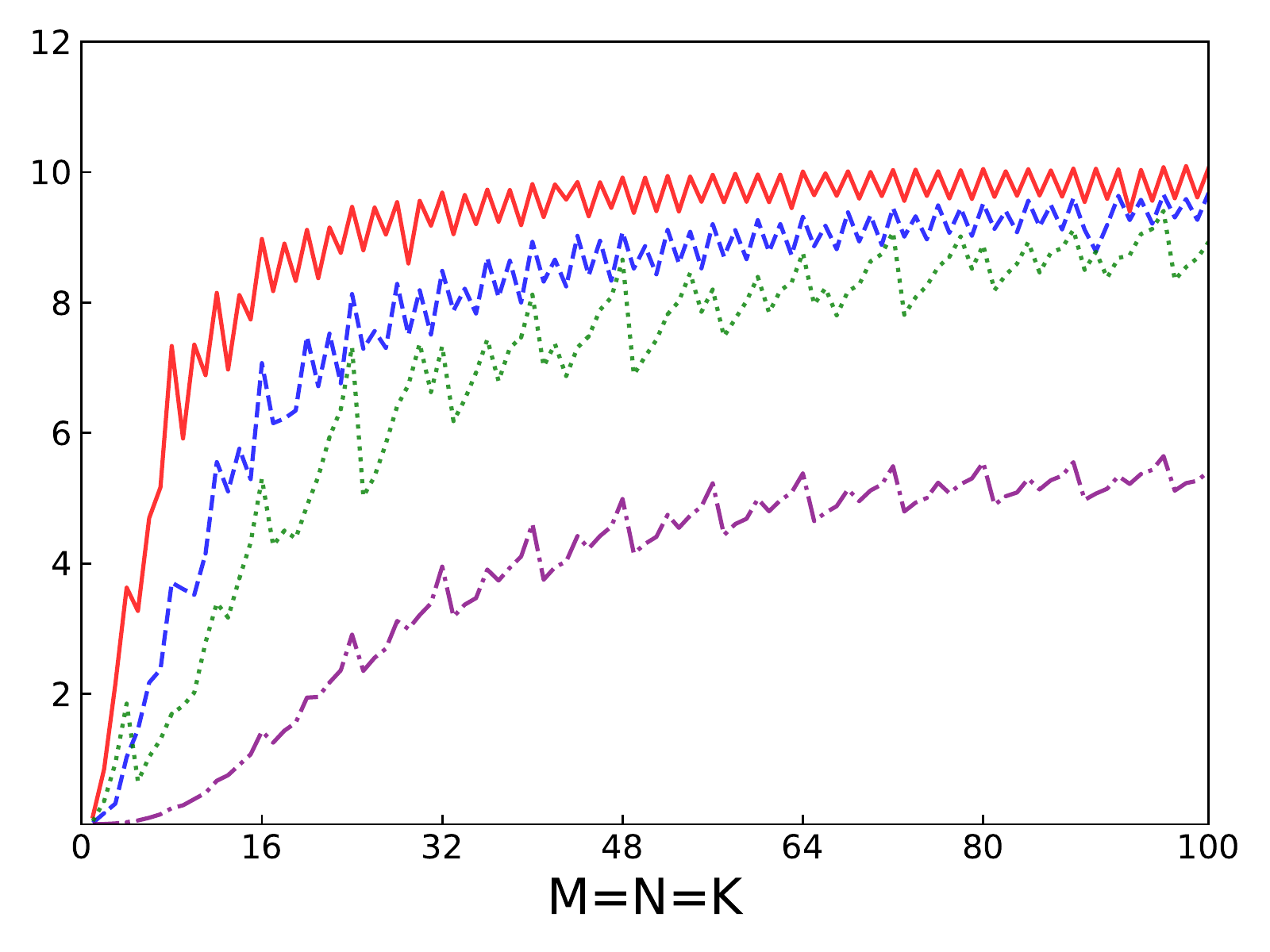}
\end{minipage}
\label{fig:dtt}
}
\centering
\caption{Performance evaluation of IAAT vs. OpenBLAS, BLIS, ARMPL for DGEMM}
\label{fig:dgemm}
\end{figure*}

\begin{figure*}[htbp]
\centering
\subfigure[NN]{
\begin{minipage}[t]{0.23\linewidth}
\centering
\includegraphics[width=4.1cm]{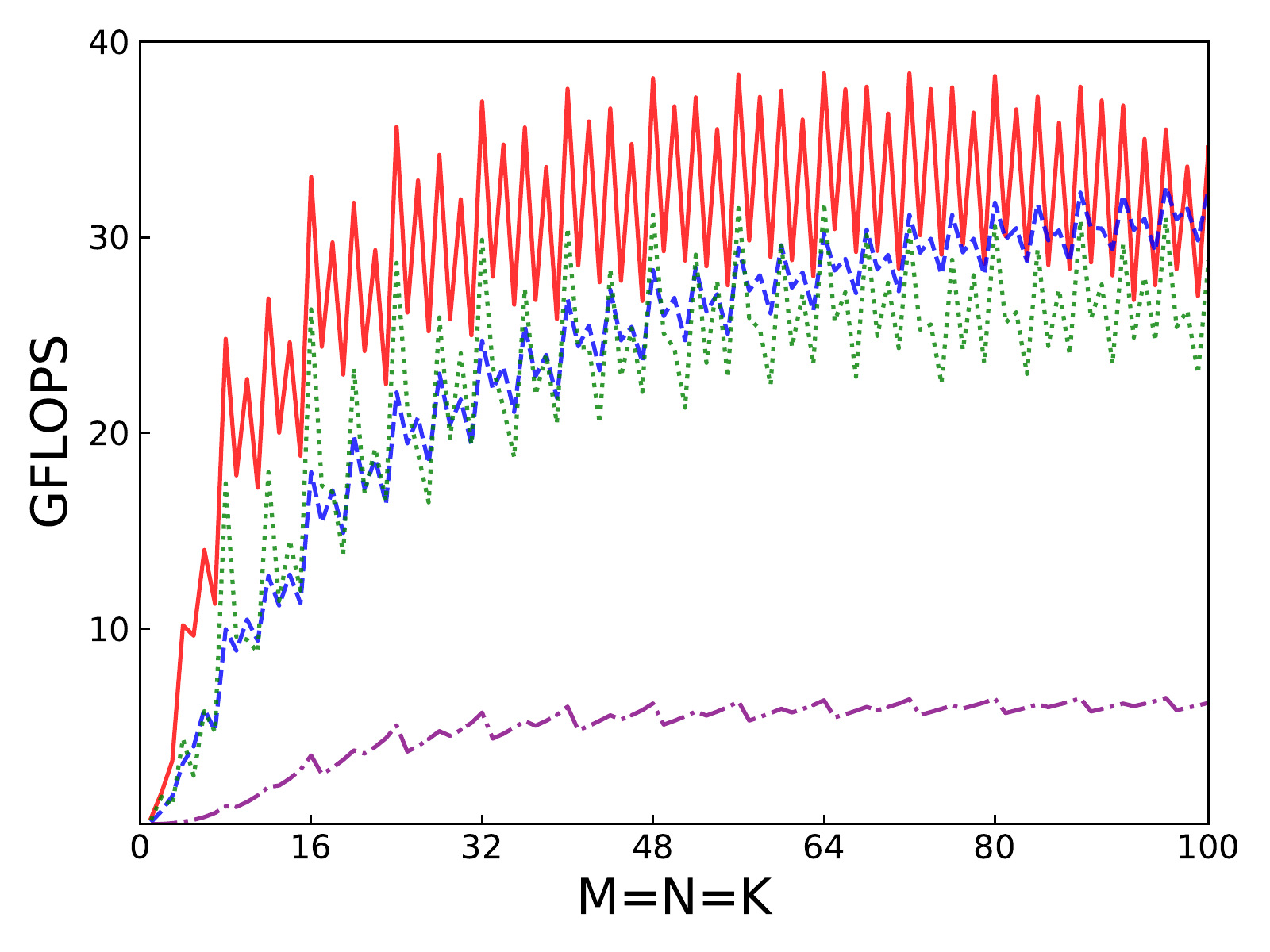}
\end{minipage}
\label{fig:cnn}
}
\hspace{-0.5cm}
\subfigure[NT]{
\begin{minipage}[t]{0.23\linewidth}
\centering
\includegraphics[width=4.1cm]{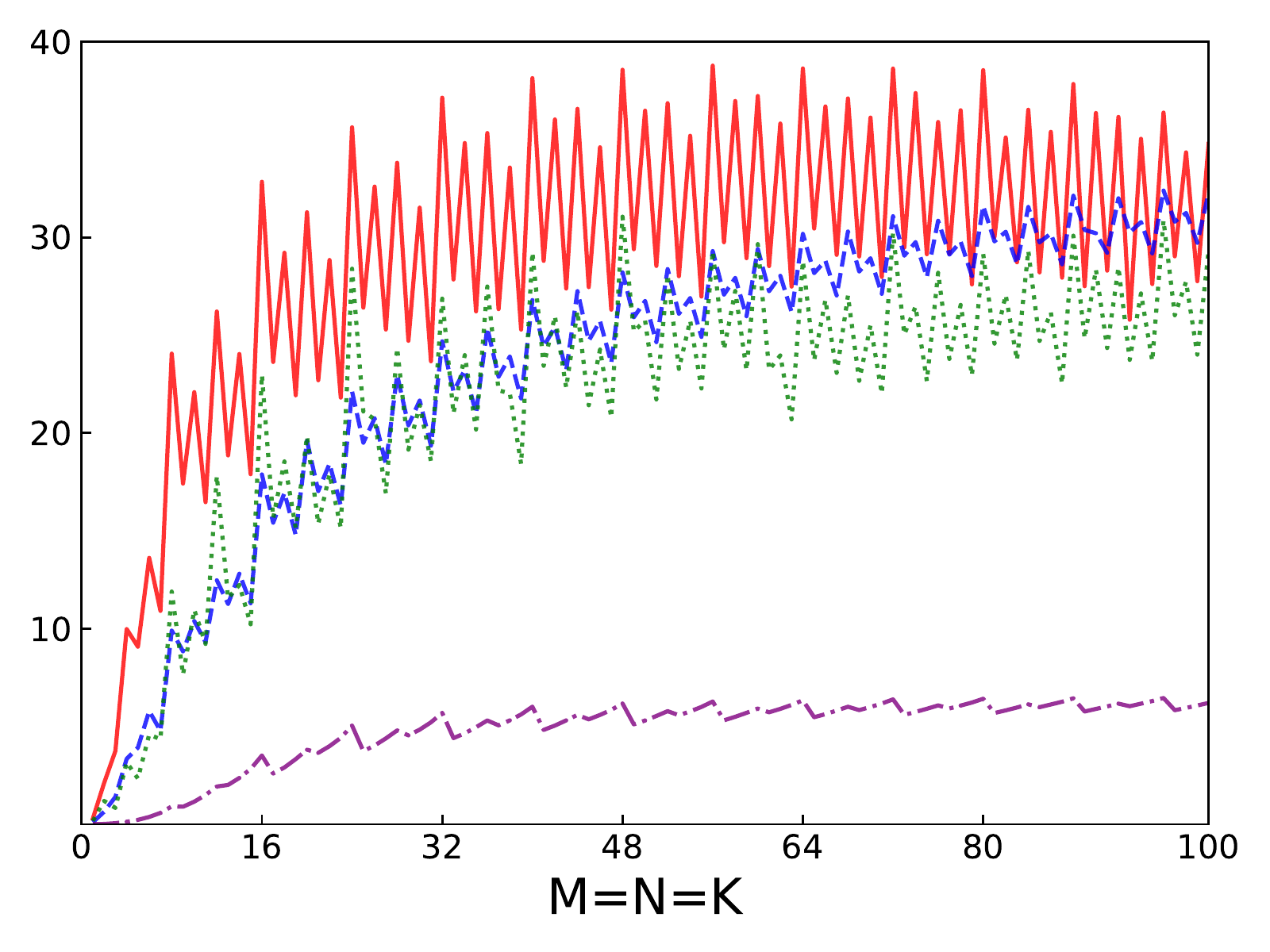}
\end{minipage}
\label{fig:cnt}
}
\hspace{-0.5cm}
\subfigure[TN]{
\begin{minipage}[t]{0.23\linewidth}
\centering
\includegraphics[width=4.1cm]{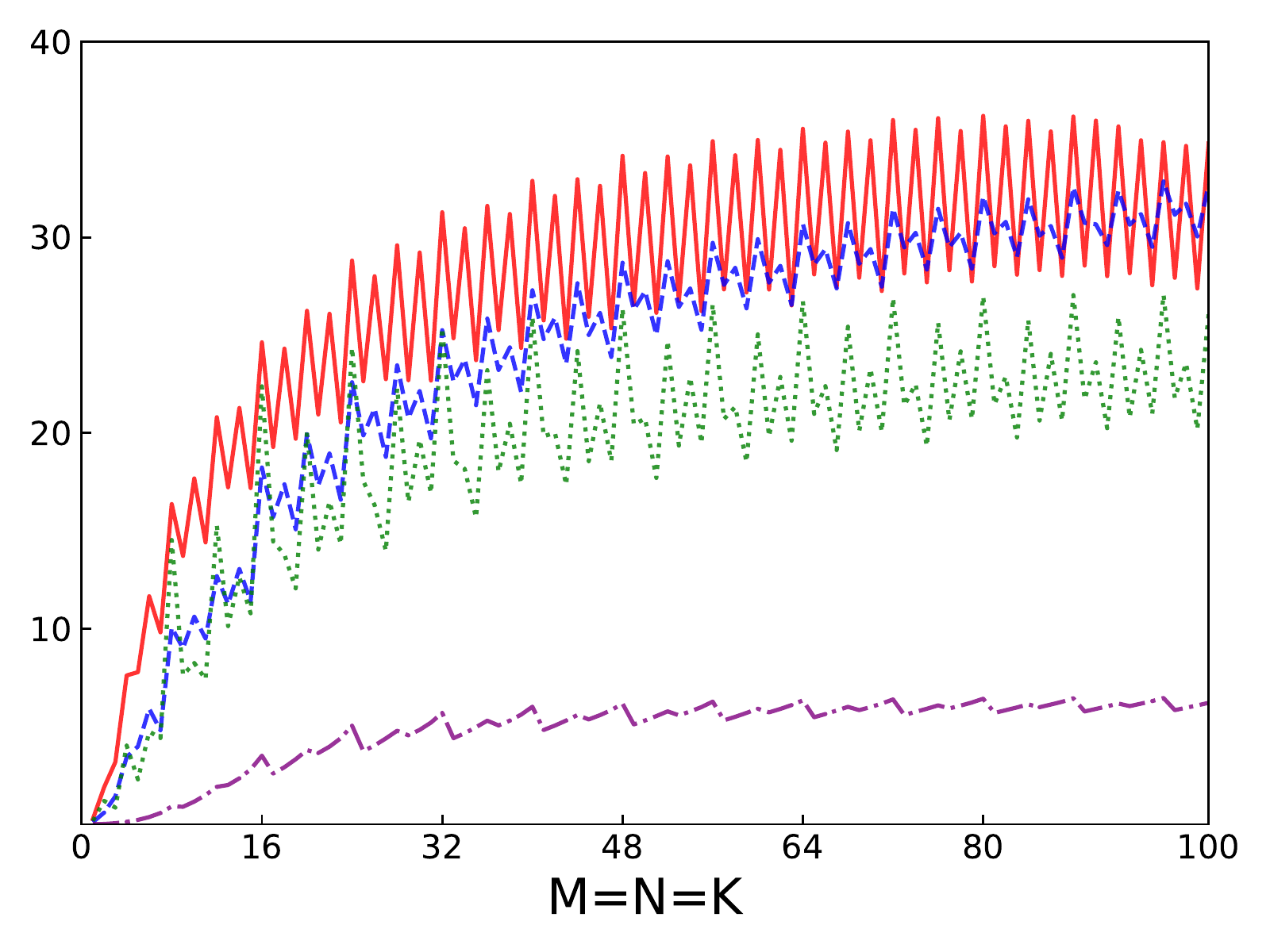}
\end{minipage}
\label{fig:ctn}
}
\hspace{-0.5cm}
\subfigure[TT]{
\begin{minipage}[t]{0.23\linewidth}
\centering
\includegraphics[width=4.1cm]{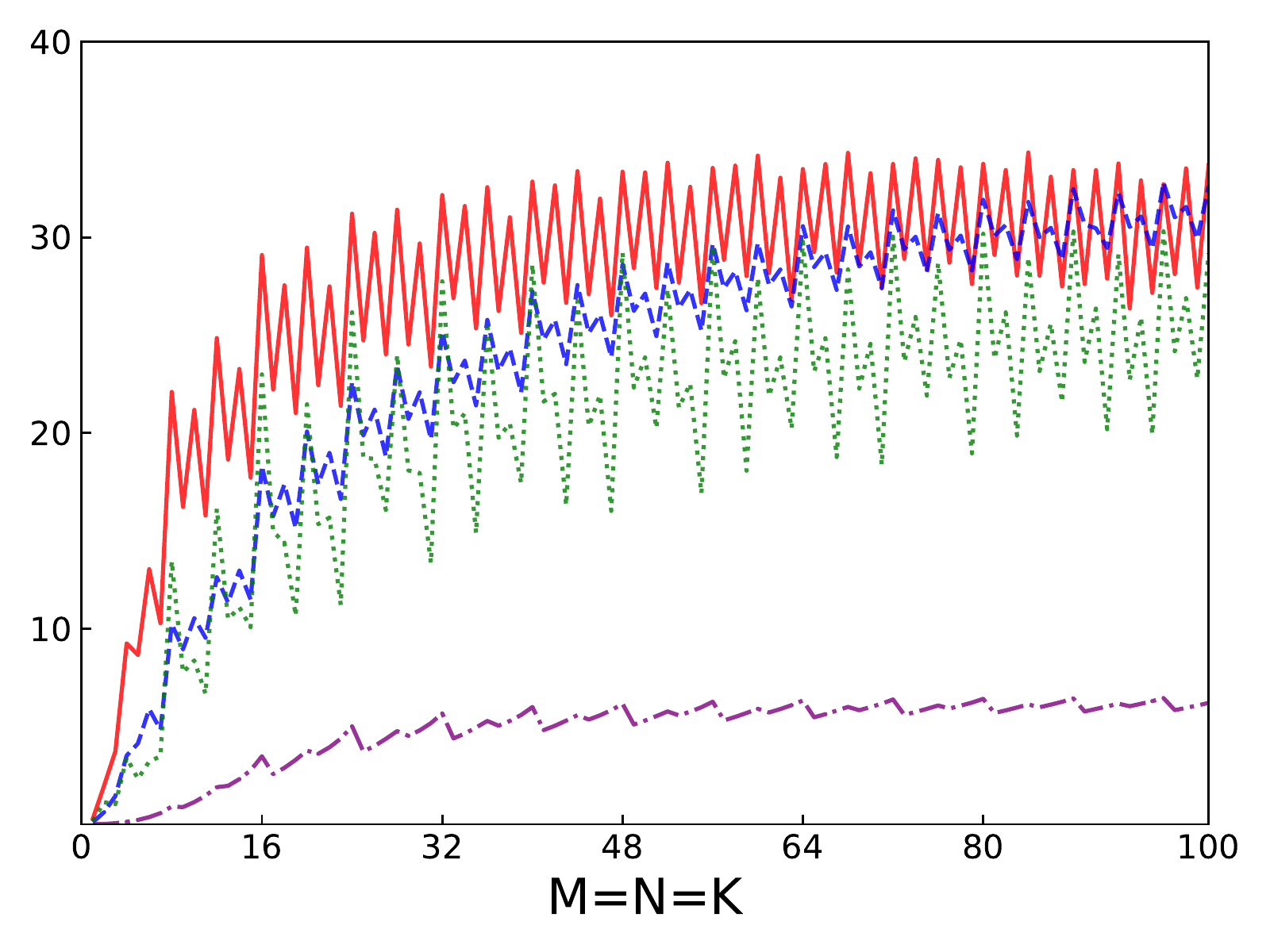}
\end{minipage}
\label{fig:ctt}
}
\centering
\caption{Performance evaluation of IAAT vs. OpenBLAS, BLIS, ARMPL for CGEMM}
\label{fig:cgemm}
\end{figure*}

\begin{figure*}[htbp]
\centering
\subfigure[NN]{
\begin{minipage}[t]{0.23\linewidth}
\centering
\includegraphics[width=4.1cm]{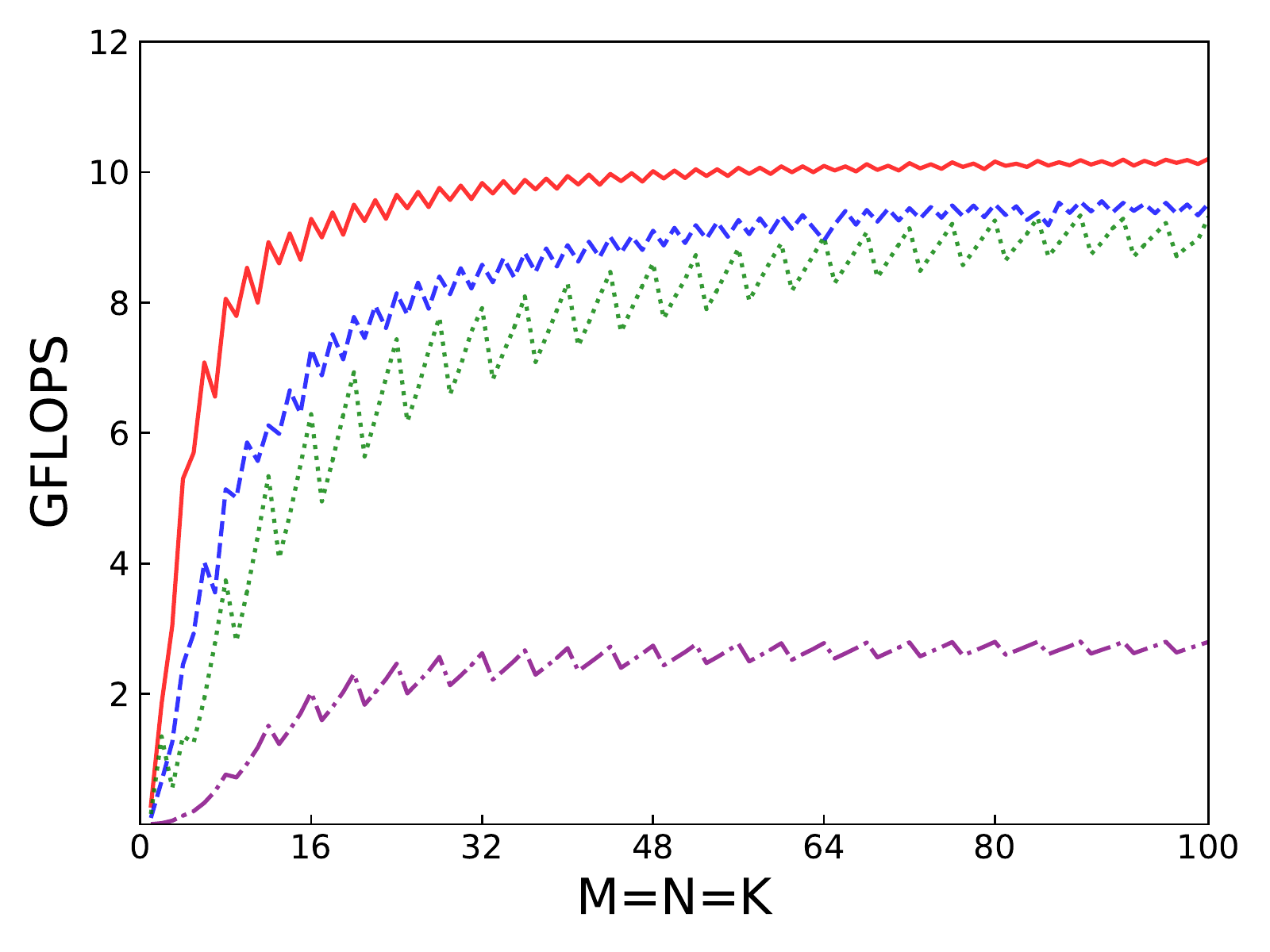}
\end{minipage}
\label{fig:znn}
}
\hspace{-0.5cm}
\subfigure[NT]{
\begin{minipage}[t]{0.23\linewidth}
\centering
\includegraphics[width=4.1cm]{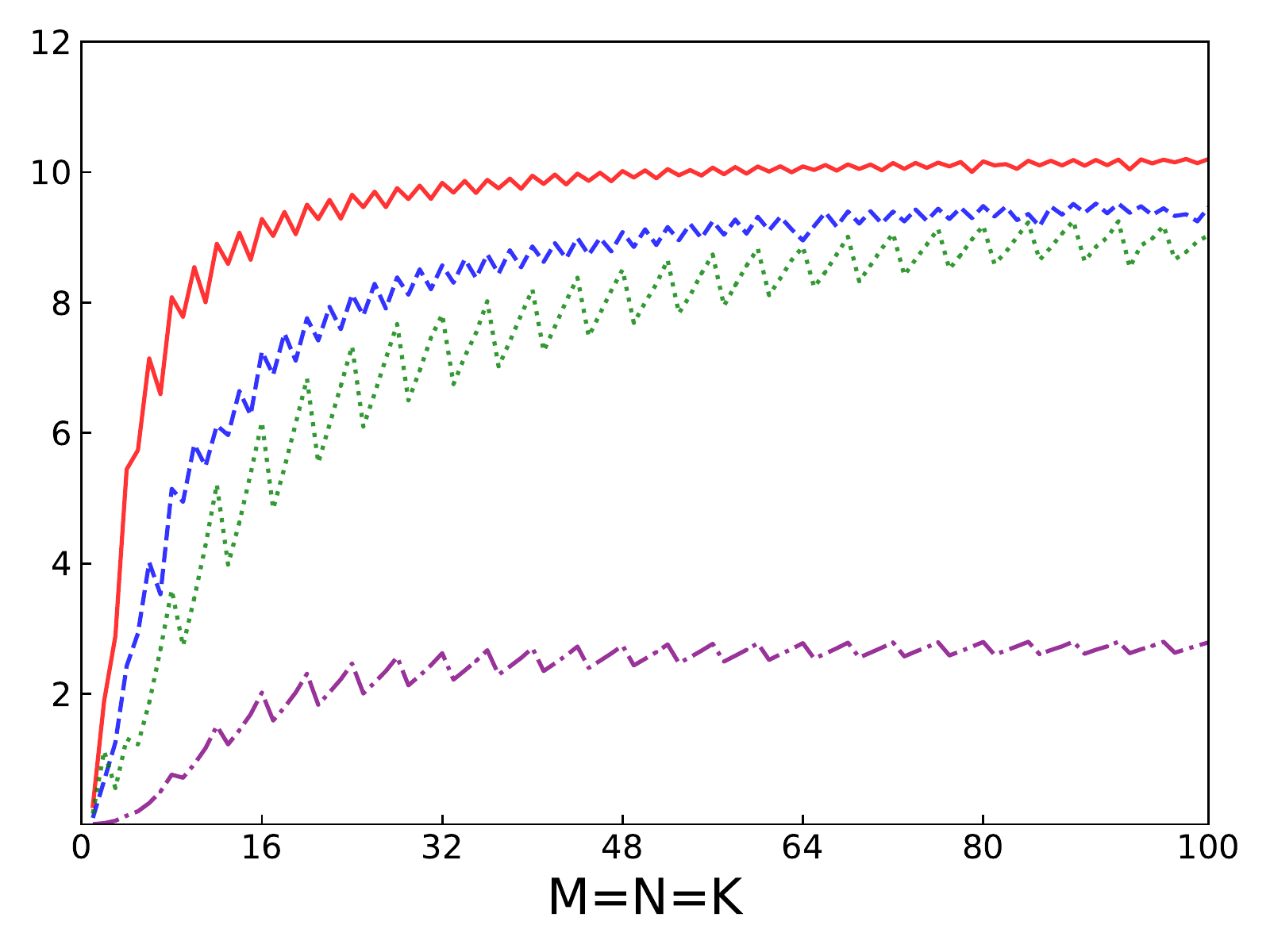}
\end{minipage}
\label{fig:znt}
}
\hspace{-0.5cm}
\subfigure[TN]{
\begin{minipage}[t]{0.23\linewidth}
\centering
\includegraphics[width=4.1cm]{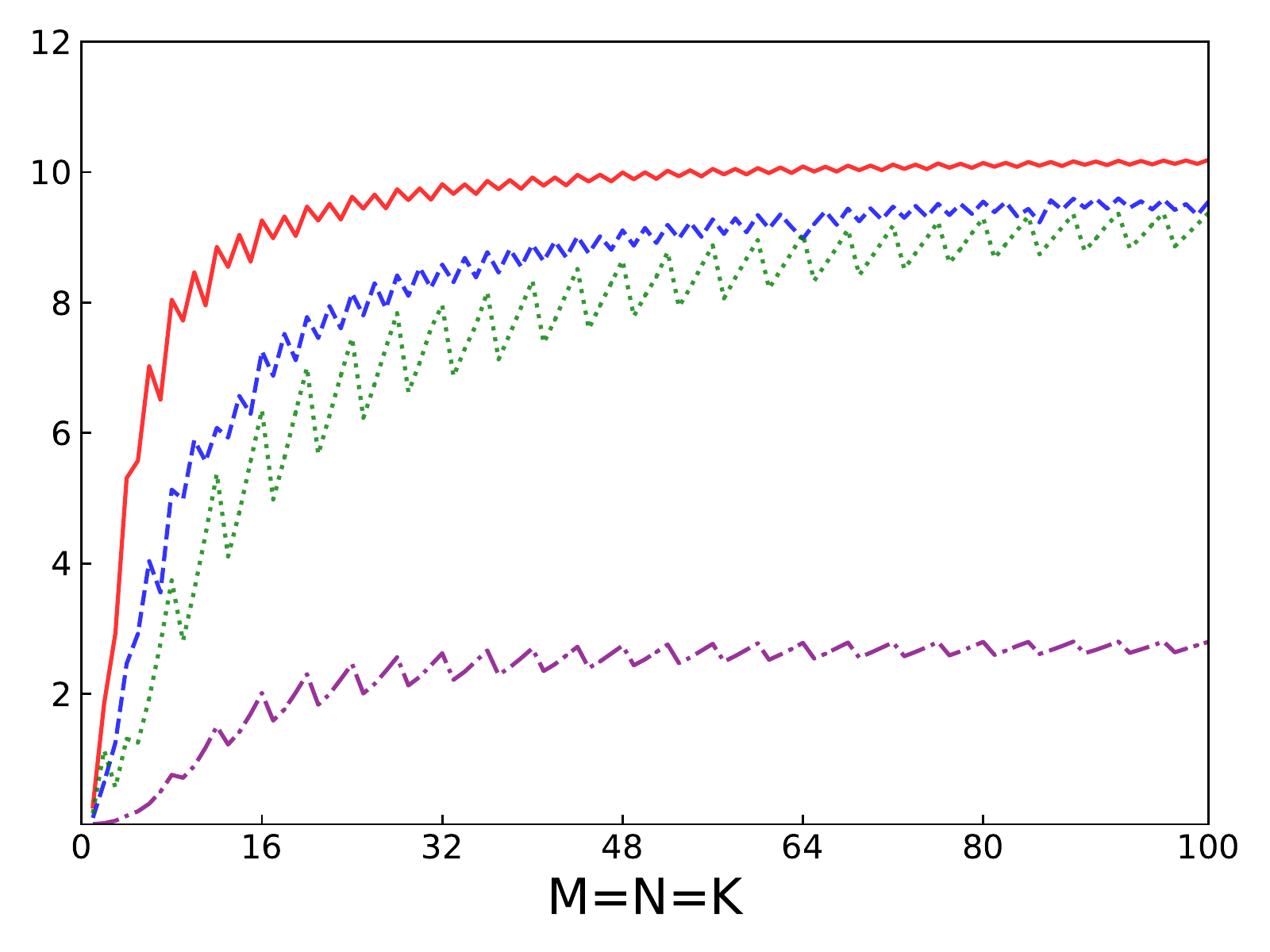}
\end{minipage}
\label{fig:ztn}
}
\hspace{-0.5cm}
\subfigure[TT]{
\begin{minipage}[t]{0.23\linewidth}
\centering
\includegraphics[width=4.1cm]{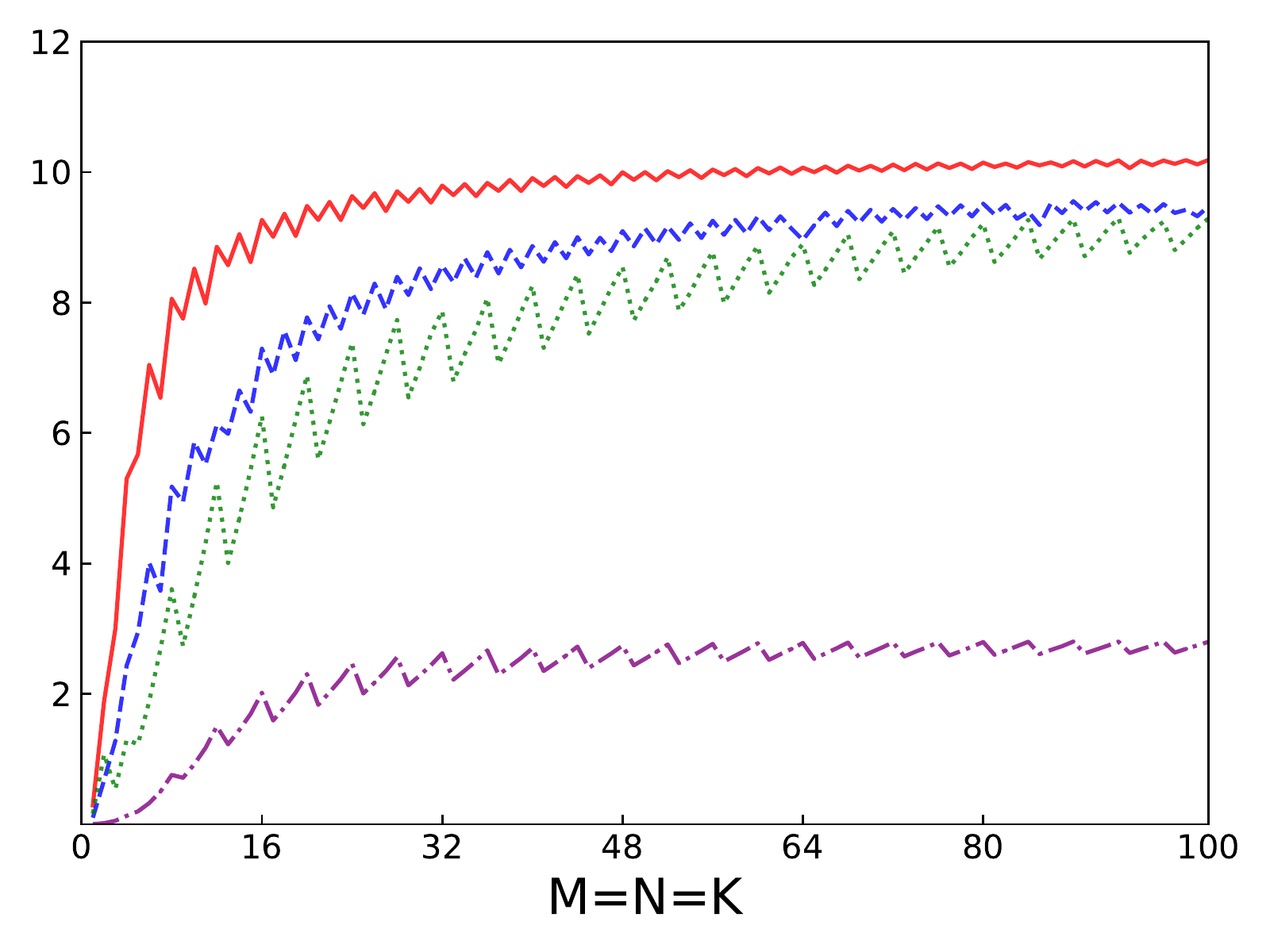}
\end{minipage}
\label{fig:ztt}
}
\centering
\caption{Performance evaluation of IAAT vs. OpenBLAS, BLIS, ARMPL for ZGEMM}
\label{fig:zgemm}
\end{figure*}

Fig.\ref{fig:sgemm} shows performances of NN, NT, TN, TT of SGEMM of IAAT, OpenBLAS, ARMPL, and BLIS. When input matrices are small, IAAT is faster than OpenBLAS, ARMPL, and BLIS for all transpositions. When $M=N=K\le 80$ and transposition is NN, as shown in Fig.\ref{fig:snn}, IAAT is on average 1.81, 2.3, and 20.17 times faster than OpenBLAS, ARMPL, and BLIS, respectively. When $M=N=K\le 80$ and transposition is NT, as shown in Fig.\ref{fig:snt}, IAAT is on average 1.81, 2.29, and 20.19 times faster than OpenBLAS, ARMPL, and BLIS, respectively. When $M=N=K\le 32$ and transposition is TN, as shown in Fig.\ref{fig:stn}, IAAT is on average 1.65 times faster than OpenBLAS. When $M=N=K>32$ and transposition is TN, as shown in Fig.\ref{fig:stn}, IAAT is only faster than OpenBLAS when sizes of input matrices are multiples of 4. However, when $M=N=K\le 100$ and transposition is TN, IAAT is faster than ARMPL and BLIS and is on average 2.15 and 11.57 times, respectively. 
When $M=N=K\le 80$ and transposition is TT, as shown in Fig.\ref{fig:stt}, IAAT is on average 1.73, 2.55, and 18.76 times faster than OpenBLAS, ARMPL, and BLIS, respectively.

Fig.\ref{fig:dgemm} shows performances of NN, NT, TN, TT of DGEMM of IAAT, OpenBLAS, ARMPL, and BLIS. When input matrices are small, IAAT is faster than OpenBLAS, ARMPL, and BLIS for all transpositions. When $M=N=K\le 80$ and transposition is NN, as shown in Fig.\ref{fig:dnn}, IAAT is on average 1.48, 1.66, and 15.0 times faster than OpenBLAS, ARMPL, and BLIS, respectively. When $M=N=K\le 80$ and transposition is NT, as shown in Fig.\ref{fig:dnt}, IAAT is on average 1.43, 1.66, and 14.56 times faster than OpenBLAS, ARMPL, and BLIS, respectively. When $M=N=K\le 80$ and transposition is TN, as shown in Fig.\ref{fig:dtn}, IAAT is on average 1.32, 1.47, and 12.78 times faster than OpenBLAS, ARMPL, and BLIS, respectively. When $M=N=K\le 80$ and transposition is TT, as shown in Fig.\ref{fig:dtt}, IAAT is on average 1.43, 1.64, and 14.54 times faster than OpenBLAS, ARMPL, and BLIS, respectively.

Fig.\ref{fig:cgemm} shows performances of NN, NT, TN, TT of CGEMM of IAAT, OpenBLAS, ARMPL, and BLIS. When input matrices are small, IAAT is faster than OpenBLAS, ARMPL, and BLIS for all transpositions. When $M=N=K\le 80$ and transposition is NN, as shown in Fig.\ref{fig:cnn}, IAAT is on average 1.31, 1.30, and 13.24 times faster than OpenBLAS, ARMPL, and BLIS, respectively. When $M=N=K\le 80$ and transposition is NT, as shown in Fig.\ref{fig:cnt}, IAAT is on average 1.37, 1.44, and 13.55 times faster than OpenBLAS, ARMPL, and BLIS, respectively. 
When $M=N=K\le 64$ and transposition is TN, as shown in Fig.\ref{fig:ctn}, IAAT is on average 1.16, 1.33, and 13.68 times faster than OpenBLAS, ARMPL, and BLIS, respectively. 
When $M=N=K\le 80$ and transposition is TT, as shown in Fig.\ref{fig:ctt}, IAAT is on average 1.27, 1.46, and 12.94 times faster than OpenBLAS, ARMPL, and BLIS, respectively.

Fig.\ref{fig:zgemm} shows performances of NN, NT, TN, TT of ZGEMM of IAAT, OpenBLAS, ARMPL, and BLIS. When input matrices are small, IAAT is faster than OpenBLAS, ARMPL, and BLIS for all transpositions. When $M=N=K\le 80$ and transposition is NN, as shown in Fig.\ref{fig:znn}, IAAT is on average 1.09, 1.3, and 9.62 times faster than OpenBLAS, ARMPL, and BLIS, respectively. When $M=N=K\le 80$ and transposition is NT, as shown in Fig.\ref{fig:znt}, IAAT is on average 1.09, 1.32, and 9.6 times faster than OpenBLAS, ARMPL, and BLIS, respectively. When $M=N=K\le 80$ and transposition is TN, as shown in Fig.\ref{fig:ztn}, IAAT is on average 1.11, 1.32, and 9.69 times faster than OpenBLAS, ARMPL, and BLIS, respectively. When $M=N=K\le 80$ and transposition is TT, as shown in Fig.\ref{fig:ztt}, IAAT is on average 1.1, 1.34, and 9.6 times faster than OpenBLAS, ARMPL, and BLIS, respectively.
ei
In addition to the above performance description, we still observe the following three phenomena.

Firstly, as shown in all Fig. \ref{fig:sgemm}, \ref{fig:dgemm}, \ref{fig:cgemm}, and \ref{fig:zgemm}, all performance curves of IAAT are very steep When input matrices are small and tend to be smooth along with increase of size. All performance curves of IAAT relate to the proportion of pack step, as shown in Fig.\ref{fig:pack}. As mentioned in Section \ref{sec:introduction}, performance improvement of small GEMM comes from removing pack steps. The greater proportion of pack step cost, the higher performance improvement. For example, when $M=N=K\le 64$, the proportion of pack step of SGEMM\_NN drops, when $M=N=K>64$, the curve of proportion is smooth. The corresponding performance curve of SGEMM\_NN, as shown in \ref{fig:snn}, rises when $M=N=K\le 64$ and stops rising when $M=N=K>64$. The others curve are the same as the curve of SGEMM\_NN.

Secondly, performance of TN transposition is not as good as other transpositions as shown in Fig.\ref{fig:sgemm}, \ref{fig:dgemm}, \ref{fig:cgemm} and \ref{fig:zgemm}. Because data is not continuous and vectorized computation is not feasible in TN transposition, register allocator has to allocate individual registers for each element of block $C_c$. Therefore, blocks of matrix C occupy too many registers, which causes kernel sizes of SGEMM\_TN smaller than other transpositions. We have to tile input matrices into smaller blocks than other transpositions. The number of loading instructions increases significantly. As the size of input matrices increases, the advantage of small GEMM for TN transposition will vanish.

Thirdly, the curves of performance of IAAT, as shown in Fig.\ref{fig:sgemm}, is wavy. The performance of four transpositions reaches wave crests when the size of input matrices is multiples of 4 and it falls into wave troughs when sizes of input matrices are not multiples of 4. Here are two reasons for this phenomenon. First, Kunpeng920 platform has two fused multiply-add(FMA) units for single but Kunpeng920 cannot fully utilize units. Kunpeng920 can issue two FMA instructions or issue one FMA instruction and one loading instruction at the same time. 
Therefore, for a kernel of any size, the lower the proportion of loading instructions, the closer the performance is to the peak performance.
When the size of the kernel is multiples of 4, the kernel can achieve better performance. When the sizes of input matrices are not multiples of 4, the tile algorithm tiles matrix into blocks, which corresponds to the low performance of kernel. Second, it is because only when the size of input matrices is multiples of 4, small GEMM can make full use of the registers. As one register can store four floats, other sizes of input matrices lead to insufficient register utilization. Therefore, the implementation of small GEMM for input matrices whose size is multiples of 4 has better performance. Similar waves occur in CGEMM as shown in Fig.\ref{fig:cgemm}, which has the same reasons as SGEMM. Besides, compared with SGEMM and CGEMM, curves of DGEMM and ZGEMM, as shown in Fig.\ref{fig:dgemm} and Fig.\ref{fig:zgemm}, are more smooth and the performance curve of DGEMM and ZGEMM reaches wave crest when the size of input matrices is multiples of 2. Because Kunpeng920 platform has one fused multiply-add(FMA) unit for double, which can be fully utilized. Besides, it is also because the size of data type of DGEMM and ZGEMM is bigger than that of SGEMM and CGEMM. DGEMM and ZGEMM can make better use of registers than SGEMM and CGEMM.

Consider the above performance results, we conclude that our implementation is faster than the others library when the size of input matrices is small enough. As shown by above performance analysis, we define small GEMM, as $\sqrt[3]{MNK}\le 80$, when transposition of input matrices is not TN, or $\sqrt[3]{MNK}\le 32$ when transposition of input matrices is TN as mentioned in Section \ref{sec:introduction}.

\section{Conclusions}
\label{sec:conclusions}
In this paper, we propose the input-aware adaptive tuning framework(IAAT) for small GEMM with two stages: the install-time stage and the run-time stage. The install-time stage auto-generates assembly kernels for ARMv8 platform and the run-time stage tiles input matrices into blocks. Finally, IAAT constructs a kernel executing plan by connecting kernels, which corresponds to the sizes of tiled blocks. As shown in the experiment, IAAT utilizes code generation and adaptive tuning to achieves near-optimal performance for small GEMM. IAAT fits the situation where computes matrix multiplication with the same size repeatedly. Our future work will focus on extending IAAT to other platforms.

\section*{Acknowledgements}
We would like to express our gratitude to all reviewer’s constructive comments for helping us polish this article. This work is supported by the National Key Research and Development Program of China under Grant Nos.2017YFB0202105, the National Natural Science Foundation of China under Grant No.61972376 and the Natural Science Foundation of Beijing under Grant No.L182053.

\bibliographystyle{IEEEtran}
\bibliography{IEEEabrv,refpaper}

\end{document}